\documentclass[journal]{IEEEtran}
\usepackage{float}
\usepackage{bbm}

\usepackage{float}
\usepackage{multirow}
\usepackage{amssymb}
\usepackage{amsmath}
\usepackage{mathrsfs}
\usepackage{amsfonts}
\usepackage{graphicx}
\usepackage{color}
\usepackage{cite}
\usepackage{amsmath}

\newcommand{\bw}{\mbox{$\bf{w}$}}
\newcommand{\bx}{\mbox{$\bf{x}$}}

\newcommand{\bA}{\mbox{$\bf{A}$}}

\newcommand{\bH}{\mbox{$\bf{H}$}}

\newcommand{\bI}{\mbox{$\bf{I}$}}
\newcommand{\bX}{\mbox{$\bf{X}$}}

\newcommand{\bW}{\mbox{$\bf{W}$}}

\newcommand{\by}{\mbox{$\bf{y}$}}
\newcommand{\bh}{\mbox{$\bf{h}$}}
\newcommand{\bn}{\mbox{$\bf{n}$}}
\newcommand{\bPi}{\mbox{$\bf{\Pi}$}}

\newcommand{\bzero}{\mbox{$\bf{0}$}}

\newcommand{\bz}{\mbox{$\bf{z}$}}

\newcommand{\HH}{\dagger}

\newtheorem{theorem}{Theorem}
\newtheorem{corollary}{Corollary}
\newtheorem{lemma}{Lemma}
\newtheorem{remark}{Remark}

\makeatletter

\newcommand{\Rmnum}[1]{\expandafter\@slowromancap\romannumeral #1@}
\IEEEoverridecommandlockouts
\makeatother

\begin{document}
\bibliographystyle{IEEEtran}
\title{Exploiting Trust Degree for Multiple-Antenna User Cooperation}

\author{Mingxiong~Zhao,
        Jong Yeol~Ryu,~\IEEEmembership{Member,~IEEE,}
        Jemin~Lee,~\IEEEmembership{Member,~IEEE,}\\
        Tony Q. S. Quek,~\IEEEmembership{Senior Member,~IEEE,} and Suili~Feng,~\IEEEmembership{Member,~IEEE}\\
\thanks{This work has been supported in part by the Key Discipline
Foundation of School of Software of Yunnan University and the
Open Foundation of Key Laboratory in Software Engineering of
Yunnan Province under Grant No.2017SE203, and the National Natural Science Foundation of China No.61640306, and the National Research Foundation of Korea (NRF) grant funded by the Korea government (MSIP) (No.2017R1C1B2009280) and Institute for Information \& communications Technology Promotion (IITP) grant funded by the Korea government (MSIP) (2014-0-00065, Resilient Cyber-Physical Systems Research), and the Guangdong Science and Technology Plan under Grant 2016A010101009.}
\thanks{M. Zhao is with the School of Software, Yunnan University, Kunming, China and also with Key Laboratory in Software Engineering of Yunnan
Province. Email: jimmyzmx@gmail.com, mx\_zhao@ynu.edu.cn.}
\thanks{J. Y. Ryu is with the Department of Information and Communication Engineering, Gyeongsang National University, Tongyeong, Korea. Email: ryujy18@gmail.com.}
\thanks{J. Lee is with the Department of Information and Communication Engineering, Daegu Gyeongbuk Insitute of Science and Technology (DGIST), Daegu, Korea. Email: jmnlee@dgist.ac.kr.}
\thanks{T. Q. S. Quek is with Singapore University of Technology and Design,
Singapore 487372 and also with the Department of Electronics Engineering,
Kyung Hee University, Yongin-si, Gyeonggi-do, 17104, Korea. E-mail:
tonyquek@sutd.edu.sg.}
\thanks{S. Feng is with the School of Electronics and Information Engineering, South China University of Technology, Guangzhou, China. Email: fengsl@scut.edu.cn.}}

\maketitle

\begin{abstract}
For a user cooperation system with multiple antennas,
we consider a trust degree based cooperation techniques to explore the influence of the trustworthiness between users on the communication systems. For the system with two communication pairs, when one communication pair achieves its quality of service (QoS) requirement, they can help the transmission
of the other communication pair according to the trust degree, which
quantifies the trustworthiness between users in the cooperation.
For given trust degree, we investigate the user cooperation strategies, which include the power allocation and precoder design for various antenna configurations.
For SISO and MISO cases, we provide the optimal power allocation and beamformer design
that maximize the expected achievable rates while guaranteeing the QoS requirement. For a SIMO case, we resort to semidefinite relaxation (SDR) technique and block coordinate update (BCU) method to solve the corresponding problem, and guarantee the rank-one solutions at each step. For a MIMO case, as MIMO is the generalization of MISO and SIMO, the similarities among their problem structures inspire us to combine the methods from MISO and SIMO together to efficiently tackle MIMO case. Simulation results show that the trust degree information has a great effect on the performance of the user cooperation in terms of the expected achievable rate, and the proposed user cooperation strategies achieve high achievable rates for given trust degree.
\end{abstract}


\section{Introduction}
The cooperative communications have been introduced to improve the communication reliability and spectral efficiency to satisfy the growing demand for higher data rates in wireless networks. Generally, there are two ways of realizing cooperation: 1) utilizing fixed relay terminals to assist the communication between dedicated sources and their corresponding destinations, 2) allowing the mobile users to help each other as a relay for reliable communication \cite{li2011diversity}. For utilizing fixed relays, the fixed relays need to be installed in the network,
which requires high infrastructure, operation and maintenance costs for the operators \cite{loodaricheh2014energy}. On the other hand, in the user cooperation, relays are the mobile users who have good channel conditions and low traffic demands, and they can help the communications without increasing the cost to the mobile operators \cite{nokleby2010user}. Furthermore, many mobile relays exist in the network, so each relay needs to assist few users only and the average power used by each mobile relay for transmission of signals is much smaller than that of the fixed relays\cite{loodaricheh2014energy}. Due to these advantages, in the cooperative communications, the user cooperation techniques have been intensively investigated \cite{nokleby2010user, loodaricheh2014energy,zhangpower,hwang2015ue,radwan2012analysis,chen2008game}.

In the user cooperation, the proximity between users enables their direct communications
via device-to-device (D2D) communication. For various communication networks,
the techniques for the cooperative D2D communications have been proposed
\cite{zhangpower,hwang2015ue,radwan2012analysis,chen2008game}.
For a cellular network, where the cellular and D2D users coexist,
the relaying scheme of D2D user to assist the downlink transmission of cellular user
was proposed in \cite{zhangpower}.
The cooperative D2D communication between the femto and macro users of a heterogeneous network (HetNet)
was proposed in \cite{hwang2015ue},
where the femto user overheard and forwarded the composite of desired and interference signals to improve the signal-to-interference ratio (SINR) of the macro user.
In \cite{radwan2012analysis}, it is shown that mobile terminals
can save the energy by exploiting the good channel quality of short range cooperation
for WiFi and WiMedia. In the existing literatures \cite{zhangpower,hwang2015ue,radwan2012analysis,chen2008game}, the user cooperation techniques are designed based on their traffic demands and corresponding qualities of physical channels.

However, different from the cooperation using the fixed relays,
in the user cooperation, the relationship between users, e.g., trust degree, may affect the performance of user cooperation\cite{coon2014modelling}. The social relationship between users can be an important motivation for participating in cooperative communications.
Users would be willing to help each other by consuming their own resources if
they have close relationship in the social domain. Otherwise, it is not sure whether users will cooperate even though they have the good channels and low traffic demands. Furthermore, some users, generally not with close relationship, may discard the data of the other user during the cooperation due to either the selfish behavior to save its own resource or the malicious purpose to disconnect the communications. Therefore, in the user cooperation, the relationship among users should be taken into account as a key design parameter for the efficient cooperative communications.

Recently, the social relationship has been actively considered in the development of communication strategies such as \cite{gong2013social,gong2014social,chen2014exploiting,
li2014social,zhang2013exploring,zhang2014social,ryu2015trust,ryu2016TWC}. With the consideration of social relationships and physical coupling among users, a social group utility maximization framework was developed to maximize the social group utility, which is a sum of individual utilities weighted by its social ties with other users in \cite{gong2013social} and \cite{gong2014social}.
The social relationship of nodes has also been considered to enhance the performance of D2D communication \cite{li2014social,chen2014exploiting,zhang2013exploring}. Specifically, social trust and social reciprocity, which were achieved by exchanging the altruistic actions among nodes, were utilized in the D2D relay selection  \cite{chen2014exploiting}, while the social-aware D2D communication architecture was proposed by exploiting social networking characteristics for system design \cite{li2014social }. In \cite{zhang2013exploring}, a traffic offloading for D2D communications was optimized for given online and offline social relations.
The trustworthiness between nodes has also been exploited for the design of efficient cooperation strategies\cite{ryu2015trust,zhang2014social, ryu2016TWC}.
In these works, the \emph{trust degree}, which quantifies
a degree of trustworthiness between nodes, was used as one of key design parameters
to develop cooperative relay frameworks for the single-input-single-output (SISO)\cite{zhang2014social} and multiple-input-single-output (MISO)\cite{ryu2015trust} systems.
In terms of communication confidentiality, it was also shown that the expected secrecy rate can be increased by exploiting the trust degree of untrustworthy node
in \cite{ryu2016TWC}.
The previous works showed that the trust information can improve the performance of the conventional systems,
which are designed based on the physical parameters only.

In this paper, motivated by the strong interests in trust degree,
we investigate the user cooperation techniques based on {trust degree between two pairs of communication users with multiple antennas, i.e., Tu$_1$-Ru$_1$ and Tu$_2$-Ru$_2$.
Different from the existing works that consider a simple system model, we consider the cooperation techniques with multiple antennas, including power allocation and precoder design.
The transmit user, Tu$_2$, helps Tu$_1$ by forwarding the information of Tu$_1$ when Tu$_2$ has good channel quality to Ru$_2$, enough to guaranteeing its own quality of service (QoS) requirement. The willingness of Tu$_2$ in helping Tu$_1$ is characterized by the trust degree, i.e.,  Tu$_2$ helps with high probability when trust degree is high.
To maximize the expected achievable rate at Ru$_1$, we jointly design the transmission strategies at Tu$_1$ and Tu$_2$ for four different antenna configurations:
1) SISO case where all users equip a single antenna as a special case,
2) MISO case where only Tu$_1$ equips multiple antennas, 3) SIMO case where only Tu$_2$ equips multiple antennas, and 4) MIMO case where both Tu$_1$ and Tu$_2$ equip multiple antennas.
For SISO case, we first present an optimal power allocation strategy at Tu$_2$, which maximizes the expected achievable rate at Ru$_1$ while guaranteeing QoS requirement at Ru$_2$. For MISO case, we provide an optimal structure of beamformer at Tu$_1$
as a linear combination of the weighted channel vectors.
Then, based on the structure, we obtain the beamformer that maximizes an approximated expected achievable rate as a function of the trust degree
and corresponding power allocation at Tu$_2$. For SIMO case, to jointly optimize the beamformers of Tu$_2$, we utilize semidefinite relaxation (SDR) technique and block coordinate update (BCU) method to solve the considered problem, and guarantee the rank-one solutions at each step. Furthermore, for MIMO case, the similarities among the problem structures of MISO, SIMO and MIMO cases inspire us to combine the design of beamformer at Tu$_1$ from MISO and the alternative algorithm from SIMO together to jointly optimize the beamformers at Tu$_1$ and Tu$_2$ to maximize the expected achievable rate at Ru$_1$.

The rest of paper is organized as follows. We first describe the trust degree and system model in Section \ref{system model}. In Section \ref{problem formulation}, based on the trust degree, the optimal transmission strategies and beamforming design in terms of the expected achievable rate are derived for four different cases. Numerical results are presented in Section \ref{simulation results}, and conclusions are drawn in Section \ref{conclusion}.

\emph{Notation:}
In this paper, lowercase and uppercase boldface letters represent
vectors and matrices, respectively.
The complex conjugate of $x$ is denoted by $\bar{x}$,
the hermitian transpose and the trace of $\mathbf{X}$ are denoted by $\mathbf{X}^\HH$ and $\text{tr}(\bX)$. $\bPi_{\bX}\triangleq \bX(\bX^\HH\bX)^{-1}\bX^\HH$ represents the orthogonal projection onto the column space of $\bX$, and $\bPi_{\bX}^\bot\triangleq \bI-\bPi_{\bX}$ denotes the orthogonal projection onto the orthogonal complement of the column space of $\bX$.
$\mathbf{X}\sim \mathcal{CN}(\mathbf{A},\mathbf{B})$ denotes
the elements of $\mathbf{X}$ that follow independent complex Gaussian
distribution with mean $\mathbf{A}$ and covariance $\mathbf{B}$.

\section{System Model}\label{system model}
We consider a system with two communication pairs: Tu$_1$-Ru$_1$ and Tu$_2$-Ru$_2$, where Tu$_1$ and Tu$_2$ are equipped with $N_1$ and $N_2$ antennas, respectively, and the receivers have a single antenna.
In our system model, if one node achieves its own QoS requirement, it can help the transmission of the other node by using residual resource. Here, we assume that the node decides whether to help the transmission or not
based on the \emph{trust degree}, which measures the trustworthiness between nodes. Without loss of generality, Tu$_2$ helps the transmission of Tu$_1$ according to the trust degree between two user pairs.
In the following subsections, we first briefly give an introduction of trust degree, and describe the system model in details.

\subsection{Trust Degree}

With the explosive growth of online social networks such as WeChat and Facebook, a growing number of people are getting involved in online social interactions, and thus, the social relationship has been studied as an important parameter to investigate how the degree of closeness of social relationship between users affects their communication strategies \cite{gong2013social,gong2014social,chen2014exploiting}.
In the communication networks, the trust degree
has been defined as a belief level that one node can put on
another node for a certain action according to previous direct
or indirect information, obtained from observations of behavior\cite{li2008future, sun2006trust}.
Hence, in the cooperative communication systems, the trust degree can be interpreted as
the degree that reveals how much a node is willing to help the communication of the other node\cite{ryu2015trust,zhang2014social}.
Similarly, in our system model, {the trust degree between Tu$_1$ and Tu$_2$}, $\alpha$, is defined by the probability that
Tu$_2$ helps the transmission of Tu$_1$ and thus, $\alpha$ is a value in range of $0\leq \alpha \leq 1$.

In the previous literatures, the trust degree has been evaluated and quantified by various ways \cite{coon2014modelling,li2008future,sun2006trust,theodorakopoulos2006trust}.
The trust degree can be evaluated by the observations of the previous behaviors of the node\cite{li2008future,sun2006trust,theodorakopoulos2006trust,Zouridaki2005,Changiz2010}.
In \cite{Zouridaki2005,Changiz2010}, the trust degree is determined using Bayesian framework.
In the Bayesian framework, the trust degree is given by the ratio of the observations of the positive behavior among total observations, where the positive behavior is that the node behaves in the predefined way of the network.
Similar to \cite{Changiz2010}, in our cooperative communication systems,
the positive behavior is defined by that Tu$_2$ helps the transmission of Tu$_1$
and hence, Tu$_1$ can estimate the trust degree based on the historical observations of the positive behavior of Tu$_2$.
The trust degree can also be updated according to new observations.
However, when the number of observations is sufficiently large, the trust degree will have ignorable change according to new observation and it will be more like a constant.
Therefore, in our system model, we assume that the trust degree is unchanged during the transmission.

In the user cooperative communications, the user
may not help the other user's transmission due to either the
selfish behavior to save its own resource or the malicious
purpose to disconnect the communication of that user. For the case of the malicious purpose,
the malicious user lets the other user know that he will help the transmission. However, the malicious user can intercept or drop the data from the other user. Therefore, for the case that the users are not trustworthy, each user designs the transmission strategy based on the trust degree.

\subsection{System Description}

In this paper, we consider the cooperative communication system,
where Tu$_2$ can help the transmission of Tu$_1$ if
its corresponding receiver Ru$_2$ achieves its QoS, as shown in Fig. \ref{F_system}.
In our system, Tu$_2$ decides whether to help the transmission of Tu$_1$
based on the trust degree, $\alpha$, which is defined as the probability that Tu$_2$ cooperates with Tu$_1$ as a relay node.
Hence, for given $\alpha$, we design the optimal transmission strategy at Tu$_1$ and
the power allocation for cooperation at Tu$_2$ to maximize an expected achievable rate
while guaranteeing the QoS requirement at Ru$_2$.
Once Tu$_2$ decides to help the transmission of Tu$_1$, Tu$_2$ determines the portion of transmission power $\beta$ for relaying information and the portion for its own data transmission.
\begin{figure}
  \centering
  \includegraphics[width=1\columnwidth]{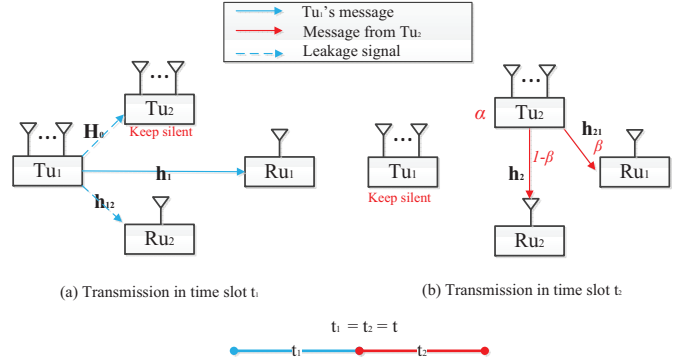}\\
   \caption{Cooperative communication system}\label{F_system}
\end{figure}
The channels from Tu$_1$ to Tu$_2$, Ru$_1$, and Ru$_2$ are defined by
$\bH_0\in \mathbb{C}^{N_2\times N_1}$, $\bh_{1}\in \mathbb{C}^{N_1\times 1}$, and $\bh_{12}\in \mathbb{C}^{N_1\times 1}$, respectively, and they follow a complex
Gaussian distribution with zero mean and covariances, $\sigma^2_{H_0}\bI_{N_2\times N_1}$,
$\sigma^2_{h_1}\bI_{N_1}$ and $\sigma^2_{h_{12}}\bI_{N_1}$, respectively.
The channels from Tu$_2$ to Ru$_1$, and Ru$_2$ are also defined by
$\bh_{21}\in \mathbb{C}^{N_2\times 1}$ and $\bh_{2}\in \mathbb{C}^{N_2\times 1}$,
which have the covariances $\sigma^2_{h_{21}}\bI_{N_2}$ and $\sigma^2_{h_2}\bI_{N_2}$, respectively. Notice that when Tu$_2$ helps the transmission of Tu$_1$, Ru$_1$ can estimate the channel from Tu$_2$, $\bh_{21}$, and then, Ru$_1$ reports the channel estimation of $\bh_{21}$ to Tu$_1$ by feedback channel.

In our system model, the data transmission operates in time-division mode, where Tu$_1$ and Tu$_2$ transmit their own data at $t_1$ and $t_2$, respectively, and $t_1$ and $t_2$ are assigned to be orthogonal with $t_1=t_2=t$. In the time slot $t_1$, Tu$_1$ transmits the information-carrying symbol $x_1$ with $E[x_1\bar{x}_1]=1$ to Ru$_1$ and during $t_1$, Ru$_2$ and Tu$_2$ can also listen to $x_1$. To efficiently transmit data, Tu$_1$ designs the transmit beamformer,
$\bw_1\in \mathbb{C}^{N_1\times 1}$, which satisfies $\bw_1^\HH\bw_1\leq P_1$ and $P_1$ is the maximum transmit budget at Tu$_1$, and uses it for transmission in $t_1$.
Hence, the received signals at Ru$_1$, Ru$_2$ and Tu$_2$ in the time slot $t_1$ are respectively given by
\begin{align}
y_{_{\mathrm{Ru}_1,1}}&=\bh_{1}^{\HH}\bw_1x_1+n_{_{\mathrm{Ru}_1,1}},\\
y_{_{\mathrm{Ru}_2,1}}&=\bh_{12}^\HH\bw_1x_1+n_{_{\mathrm{Ru}_2,1}},\\
\by_{_{\mathrm{Tu}_2}}&=\bH_0\bw_1x_1+\bn_{_{\mathrm{Tu}_2}},
\end{align}
where $n_{_{\mathrm{Ru}_1,1}}$ and $\ n_{_{\mathrm{Ru}_2,1}}$ represent the additive white Gaussian noise (AWGN) associated with Ru$_1$ and Ru$_2$, which follow the complex Gaussian distribution with zero mean and variance $\sigma^2$ (i.e., ${\cal{CN}}(0,\sigma^2)$), and $\bn_{_{\mathrm{Tu}_2}}\in \mathbb{C}^{N_2\times 1}$ is the AWGN at Tu$_2$ with ${\cal{CN}}(\bzero,\sigma^2\bI_{_{N_2}})$.
We assume that if Ru$_2$ can decode the data from the received signal, $y_{_{\mathrm{Ru}_2,1}}$, in the time slot $t_1$, Ru$_2$ can use it as side information for further improving the performance by interference cancellation in $t_2$.

In the time slot $t_2$, Tu$_2$ transmits $x_2$ with $E[x_2\bar{x}_2]=1$ to Ru$_2$.
Here, Tu$_2$ has its traffic demand $Q$ and if Ru$_2$ achieves $Q$ by the part of total power budget,
Tu$_2$ can help the transmission of Tu$_1$ using residual power.
According to the trust degree, Tu$_2$ relays the data of Tu$_1$ to Ru$_1$ via decode-and-forward (DF) based relaying with probability $\alpha$ and otherwise, Tu$_2$ transmits its own data only. Thus, the transmitted signal from Tu$_2$ can be represented by
\begin{equation}
\bx_{_{\mathrm{Tu}_2}}=\bw_{21}x_1+\bw_{22}x_2,
\end{equation}
where $\bw_{21}$ and $\bw_{22}$ are the transmit beamformers at Tu$_2$ for $x_1$ and $x_2$, respectively,
and they are designed to satisfy the power constraint as
\begin{equation}
\bw_{21}^\HH\bw_{21}+\bw_{22}^\HH\bw_{22}\leq P_2,
\end{equation}
where $P_2$ is the maximum transmit budget at Tu$_2$
and $\bw_{21}^\HH\bw_{21}=\beta P_2$.
If Tu$_2$ does not help Tu$_1$ with probability $1-\alpha$,
Tu$_2$ does not allocate the power for $x_1$ such as $\beta=0$.
Therefore, the received signals at Ru$_1$ and Ru$_2$ in the time slot $t_2$ are respectively given by
\begin{align}
\!\!&\!y_{_{\mathrm{Ru}_1,2}}\!\!=\!\bh_{21}^\HH\bx_{_{\mathrm{Tu}_2}}\!\!\!+\!n_{_{\mathrm{Ru}_1,2}}\!\!=\!\!\bh_{21}^\HH\bw_{21}x_1\!+\!\bh_{21}^\HH\bw_{22}x_2\!\!+\!n_{_{\mathrm{Ru}_1,2}},\\
\!\!&\!y_{_{\mathrm{Ru}_2,2}}\!\!=\!\bh_{2}^\HH\bx_{_{\mathrm{Tu}_2}}\!\!\!+\!n_{_{\mathrm{Ru}_2,2}}\!\!=\!\bh_{2}^\HH\bw_{22}x_2\!+\!\bh_{2}^\HH\bw_{21}x_1\!\!+\!n_{_{\mathrm{Ru}_2,2}},
\end{align}
where $n_{_{\mathrm{Ru}_1,2}}$ and $\ n_{_{\mathrm{Ru}_2,2}}$ are the AWGN at Ru$_1$ and Ru$_2$ with ${\cal{CN}}(0,\sigma^2)$ in $t_2$, respectively.
If Ru$_2$ successfully decodes the data from Tu$_1$ in the time slot $t_1$,
Ru$_2$ can subtract it from the received signal in $t_2$ by applying the successive interference cancellation (SIC).
For this case, after applying SIC, the received signal at Ru$_2$ in $t_2$ can be rewritten as $y_{_{\mathrm{Ru}_2,2}}^{\text{SIC}}=\bh_{2}^\HH\bw_{22}x_2+n_{_{\mathrm{Ru}_2,2}}$. Otherwise, Ru$_2$ has to decode $x_2$ by treating the signal related to $x_1$ as the noise.

In this paper, we only consider the case that Ru$_2$ can always achieve its QoS requirement, $Q$, for given power budget, $P_2$, and hence, the QoS requirement at Ru$_2$ is given in the range of $0\leq Q \leq Q^{\max}$, where the maximum QoS requirement is $Q^{\max}=\frac{1}{2}\log_2\left(1+\frac{P_2\|\bh_{2}\|^2}{\sigma^2}\right)$, which is achieved by the maximal-ratio transmission (MRT) at Tu$_2$, $\bw_2^\mathrm{mrt}=\sqrt{P_2}\frac{\mathbf{h}_{2}}{\|\mathbf{h}_{2}\|}$ with maximum power $P_2$, and $\frac{1}{2}$ is from the fact that the transmission takes place in two time slots.

\section{User Cooperation based on Trust Degree}\label{problem formulation}
In this section, for given trust degree, we provide the cooperation strategy, which includes the transmission beamformer at Tu$_1$ and the power allocation for cooperation at Tu$_2$, to maximize the expected achievable rate at Ru$_1$ while guaranteeing QoS requirement at Ru$_2$. We derive the optimal transmission strategies for three cases: 1) SISO case ($N_1=N_2=1$), 2) MISO case ($N_1\geq2$, $N_2=1$), 3) SIMO case ($N_1=1$, $N_2\geq 2$), and 4) MIMO case ($N_1\geq 2,\ N_2\geq 2$). We first define the event of cooperation as $\mathcal{E}$, where
$\mathcal{E}=1$ and $\mathcal{E}=0$ stand for the events that Tu$_2$ helps and does not help the transmission, respectively. Thus, $\mathcal{E}$ is a Bernoulli random variable with $Pr[\mathcal{E}=1]=\alpha$
and $Pr[\mathcal{E}=0]=1-\alpha$. As a performance metric, for given trust degree $\alpha$, we use the expected achievable rate with respect to the possible cooperation events, defined as
\begin{align}
R_{_{\mathrm{Ru}_1}}=\underset{\mathcal{E}}{\mathbb{E}}\left\{\tilde{R}_{_{\mathrm{Ru}_1}}\right\},
\end{align}
where $\tilde{R}_{_{\mathrm{Ru}_1}}$ is an achievable rate at Ru$_1$.

\subsection{SISO case ($N_1=N_2=1$)}
We first consider a simple SISO case that Tu$_1$ and Tu$_2$ have a single antenna ($N_1=N_2=1$).
We define the gains of all channels as
\begin{equation}
g_0\!=\!|h_0|^2,\!g_1\!=\!|h_1|^2,\!g_2\!=\!|h_2|^2,\!g_{12}\!=\!|h_{12}|^2,\! g_{21}\!=\!|h_{21}|^2.\!\!\!\!\!\!
\end{equation}

When the channel condition between Tu$_1$ and Tu$_2$ is worse than the direct channel from Tu$_1$ to Ru$_1$ (i.e.
$g_0\leq g_1$), the cooperation of Tu$_2$ cannot improve the achievable rate at Ru$_1$ due to DF relaying constraint\cite{laneman2004cooperative}.
Thus, in this case, the achievable rate at Ru$_1$ is achieved by the direct transmission from Tu$_1$.
Therefore, since Tu$_2$ helps the transmission of Tu$_1$ with probability $\alpha$,
the expected achievable rate at Ru$_1$ is given by
\begin{align}\label{Ru1}
R_{_{\mathrm{Ru}_1}}(\beta)=\left\{\begin{array}{ll}
                       \bar{R}_{_{\mathrm{Ru}_1}}(\beta), & \mathrm{if}\  g_0> g_1, \\
                       \frac{1}{2}\log_2(1+\rho_1g_1), & \mathrm{otherwise},
                     \end{array}\right.
\end{align}
where $\bar{R}_{_{\mathrm{Ru}_1}}(\beta)$ is given by
\begin{align}
\bar{R}_{_{\mathrm{Ru}_1}}\!(\beta)=&\frac{\alpha}{2}\min\left[\log_2\!\left(1\!+\!\rho_1g_1\!+\!
\frac{\beta\rho_2g_{21}}{(1-\beta)\rho_2 g_{21}\!+\!1}\right)\!,\right.\nonumber\\
&\left.\log_2(1\!+\!\rho_1g_0)\vphantom{\log_2\left(1+\rho_1g_1+\frac{\beta\rho_2g_{21}}{(1-\beta)\rho_2 g_{21}+1}\right)}\right]\!+\!\frac{1\!-\!\alpha}{2}\!\log_2\!\left(1\!+\!\rho_1g_1\right),\label{R1}
\end{align}
where $\rho_1=\frac{P_1}{\sigma^2}$ and $\rho_2=\frac{P_2}{\sigma^2}$. From the observations of \eqref{Ru1} and \eqref{R1} in SISO case, we deduce that $R_{_{\mathrm{Ru}_1}}$ is an increasing function of $\rho_1$. Therefore, as $P_1$ grows, the value of $R_{_{\mathrm{Ru}_1}}$ is increasing, and the maximum transmit power $P_1$ is always optimal. The first term of \eqref{R1} denotes the achievable rate at Ru$_1$ when Tu$_2$ helps transmission of Tu$_1$ with probability $\alpha$, and it is bounded by the minimum of achievable rates at Tu$_2$ and Ru$_1$ due to the constraint of DF relaying \cite{laneman2004cooperative}. The second term of \eqref{R1} represents the achievable rate achieved by direct transmission from Tu$_1$ to Ru$_1$ when Tu$_2$ does not help transmission of Tu$_1$ with probability $(1-\alpha)$.
In the SISO case, we can see that $\beta$ can be determined independently from $\alpha$.
The power allocation $\beta$ has to be jointly determined with the transmit strategy at Tu$_1$,
which is related to $\alpha$.
However, in SISO case, Tu$_1$ does not design the transmit beamformer and hence,
$\beta$ can be determined independently with $\alpha$ to maximize the rate achieved when Tu$_2$ cooperates with Tu$_1$.

We define the part related with $\beta$ in \eqref{R1} as $Q_{_{\mathrm{Tu}_1}}\!(\beta)$, given by
\begin{align}\label{Q_Tu}
Q_{_{\mathrm{Tu}_1}}\!(\beta)=&\frac{1}{2}\min\left[\log_2\left(1+\rho_1g_1+\frac{\beta\rho_2g_{21}}{(1-\beta)\rho_2 g_{21}+1}\right)\!,\right.\nonumber\\
&\left.\log_2\left(1+\rho_1g_0\right)\vphantom{\log_2\left(1+\rho_1g_1+\frac{\beta\rho_2g_{21}}{(1-\beta)\rho_2 g_{21}+1}\right)}\right].
\end{align}
According to $\beta$, $Q_{_{\mathrm{Tu}_1}}\!(\beta)$ can be rewritten by
\begin{equation}\label{min}
Q_{_{\mathrm{Tu}_1}}\!(\beta)\!\!=\!\!\left\{\!\!\begin{array}{ll}
                                   \frac{1}{2}\!\log_2(1\!\!+\!\!\rho_1g_0), ~~~~~~~~~~~~~~\text{if}\ \!\beta_{0}\!\leq\!\beta\!\leq\!1,\\
                                   \frac{1}{2}\!\log_2\!\left(\!1\!\!+\!\!\rho_1g_1\!\!+\!\!\frac{\beta\rho_2g_{21}}{(1\!-\!\beta)\rho_2 g_{21}\!+\!1}\!\right),\mathrm{otherwise},
                                 \end{array}\right.
\end{equation}
where $\beta_{0}=1-\frac{\rho_2g_{21}-\rho_1( g_0- g_1)}{\rho_2 g_{21}\left\{1+\rho_1(g_0-g_1)\right\}}$.

For the case that Tu$_2$ helps the transmission of Tu$_1$\footnote{Since the QoS requirement, $Q$, is always satisfied when Tu$_2$ does not help the transmission of Tu$_1$,
we only consider the case that Tu$_2$ helps the transmission of Tu$_1$ to design the transmission strategy.}, if the receiving rate at Ru$_2$ in $t_1$, $R_{12}$, is greater than $Q_{_{\mathrm{Tu}_1}}(\beta)$, i.e., $R_{12}=\frac{1}{2}\log_2(1+\rho_1g_{12})\geq Q_{_{\mathrm{Tu}_1}}(\beta)$, Ru$_2$ can decode $x_1$ in $t_1$ and hence, Ru$_2$ can apply SIC to eliminate the effect of $x_1$ in its received signal in $t_2$.
Therefore, the achievable rate at Ru$_2$ is given by
\begin{align}\label{R2}
R_{_{Ru_2}}(\beta)\!\!=\!\!\left\{\begin{array}{ll}
                       R_{_{\mathrm{Ru}_2}}^\mathrm{SIC}(\beta), &\!\!\!\!\!\! \mathrm{if}\ R_{12}\!\!\geq\!\! Q_{_{\mathrm{Tu}_1}}\!(\beta),g_0\!\!>\!\!g_1, \\
                       R_{_{\mathrm{Ru}_2}}^\mathrm{NSIC}(\beta), &\!\!\!\!\!\! \mathrm{if}\ R_{12}\!\!<\!\!Q_{_{\mathrm{Tu}_1}}\!(\beta),g_0\!\!>\!\!g_1, \\
                       \frac{1}{2}\!\log_2(\!1\!\!+\!\rho_2g_2\!), &\!\!\!\!\!\! \mathrm{otherwise},
                     \end{array}\right.
\end{align}
where $R_{_{\mathrm{Ru}_2}}^\mathrm{SIC}(\beta)$ and $R_{_{\mathrm{Ru}_2}}^\mathrm{NSIC}(\beta)$ are given, respectively, by
\begin{align}
\left.\begin{array}{ll}
R_{_{\mathrm{Ru}_2}}^\mathrm{SIC}(\beta)=\frac{1}{2}\log_2\left(1+(1-\beta)\rho_2g_2\right),\\
R_{_{\mathrm{Ru}_2}}^\mathrm{NSIC}(\beta)=\frac{1}{2}\log_2\left(1+\frac{(1-\beta)\rho_2 g_2}{\beta\rho_2 g_2+1}\right).
\end{array}\right.\label{R2-NSIC-SIC}
\end{align}

For the SISO case, the optimal power allocation of Tu$_2$ that maximizes the expected achievable rate at Ru$_1$
while guaranteeing QoS requirement at Ru$_2$ is obtained by the following problem
\begin{subequations}\label{OP1}
\begin{align}
\mathbf{P1}:~\max_{0\leq \beta \leq 1}~& R_{_{\mathrm{Ru}_1}}\!(\beta)\\
\mathrm{s.t.} ~& R_{_{\mathrm{Ru}_2}}\!(\beta)\geq Q,\label{OP1-C1}
\end{align}
\end{subequations}
where $R_{_{\mathrm{Ru}_1}}\!(\beta)$ is given in \eqref{Ru1} and $Q\in [0,Q^{\max}]$ is QoS of Ru$_2$, where $Q^{\max}=\frac{1}{2}\log_2(1+\rho_2 g_2)$.

For given channel conditions and QoS at Ru$_2$, we obtain the optimal power allocation of Tu$_2$ for cooperative transmission in the following theorem.

\begin{theorem}\label{theorem}
For given channels and QoS requirement, $Q$,
the optimal power allocation of Tu$_2$ for cooperative transmission that maximizes
the expected achievable rate at Ru$_1$
is obtained by
\begin{align}\label{beta}
\beta^\star=\left\{\begin{array}{ll}
                                    \beta_{Q_1},&\mathrm{if}~\left\{\begin{array}{l}
                                    \! g_{12}\geq g_0,~Q\leq r_1, \\
                                    \! g_{12}\geq g_1,~Q\geq\max{(r_1,r_2)},\\
                                    \end{array}\right.\\
                                    \tilde{\beta}_{1},&\mathrm{if}~\ g_0>g_{12}\geq g_1,~r_2\geq Q> r_3,\\
                                    \beta_{Q_2},&\mathrm{otherwise},\\
                                 \end{array}\right.
\end{align}
where $\beta_{Q_1}$, $\beta_{Q_2}$, and $\tilde{\beta}_{1}$ are given, respectively, by
\begin{align}
&\beta_{Q_1}=1-\frac{4^Q-1}{\rho_2g_2},\label{betaQ1}\\
&\beta_{Q_2}=\left(1-\frac{4^Q-1}{\rho_2g_2}\right)4^{-Q},\label{betaQ2}\\
&\tilde{\beta}_{1}=1-\frac{\rho_2g_{21}-\rho_1( g_{12}- g_1)}
{\rho_2 g_{21}\left\{1+\rho_1(g_{12}-g_1)\right\}}\label{betatilde},
\end{align}
and $r_1$, $r_2$ and $r_3$ are given by
\begin{align}
\left.\begin{array}{ll}
  r_1=\Big[\frac{1}{2}\log_2\Big(1\!+\!(1\!-\!{\beta}_{0})\rho_2g_2\Big)\Big]^{+},\\
  r_2=\left[\frac{1}{2}\log_2\left(1\!+\!(1\!-\!\tilde{\beta}_{1})\rho_2g_2\right)\right]^{+},\\
  r_3=\frac{1}{2}\log_2\left(\frac{1\!+\!\rho_2g_2}{1\!+\!\rho_2g_2 \tilde{\beta}_{1}}\right).
  \end{array}\right.\label{rs}
\end{align}
\end{theorem}
\begin{IEEEproof}
The proof is presented in Appendix \ref{app1}.
\end{IEEEproof}

\begin{remark}
From Theorem \ref{theorem}, we observe that the optimal $\beta$ that maximizes the expected achievable rate at Ru$_1$
while guaranteeing QoS of Ru$_2$ is mainly determined by channel quality from Tu$_1$ to Ru$_2$, $g_{12}$ and QoS of Ru$_2$, $Q$.
Since Tu$_2$ can assign more power for cooperation when Ru$_2$ can apply SIC than that when Ru$_2$ cannot apply SIC, the optimal power allocation is mainly determined by the parameters that decide whether SIC is applicable at Ru$_2$, $g_{12}$ and $Q$.

For the case that the channel quality from Tu$_1$ to Ru$_2$ is good such as $g_{12}\geq g_0$ and $g_{12}\geq g_1$, since we have $r_1\geq r_2$ from $\beta_{0}\leq \tilde{\beta}_{1}$, the optimal power allocation for cooperation is given by $\beta^\star=\beta_{Q_1}$ for all $Q$.
In this case, for all $Q$, Ru$_2$ can apply SIC due to good channel quality between Tu$_1$ and Ru$_2$.
Contrarily, when the channel quality from Tu$_1$ to Ru$_2$ is poor such as $g_0>g_{12}$ and $g_1>g_{12}$, Ru$_2$ cannot apply SIC for all $Q$ and hence, the Tu$_2$ allocates
the minimum power for cooperation as $\beta^\star=\beta_{Q_2}$.

On the other hand, for the moderate quality of $g_{12}$ as $g_0>g_{12}\geq g_1$, the power allocation is also determined according to QoS requirement, $Q$.
In this case, since Ru$_2$ cannot always apply SIC due to $g_{12}$ and $Q$, $\beta$ has to be controlled to apply SIC.
First, for the case that QoS requirement is small such as $Q\leq r_3$, since Tu$_2$ can allocate large power for cooperation without SIC as $\beta^\star=\beta_{Q_2}$, which is a decreasing function with $Q$,
SIC by reducing $\beta$ is not beneficial.
For moderate QoS as $r_{2}\geq Q > r_{3}$, Tu$_2$ controls $\beta$ as $\beta^\star=\tilde{\beta}_{1}$, which is a constant for given channels, to apply SIC. In the moderate $Q$, the power for cooperation is not decreased
even if $Q$ increases.
However, when QoS requirement is high as $Q\geq r_2$, the power allocation for cooperation is decreased according to $Q$ as $\beta^\star=\beta_{Q_1}$ to guarantee high QoS by applying SIC.
\end{remark}

\subsection{MISO case ($N_1\geq 2$ and $N_2=1$)}\label{MISO-B}
In this subsection, we consider the MISO case, where Tu$_1$ equips with multiple antennas (i.e., $N_1\geq2$) but Tu$_2$ has a single antenna (i.e., $N_2=1$).
Hence, in the MISO case, we jointly design the transmit beamformer at Tu$_1$, $\bw_1$, and
the power allocation for cooperation at Tu$_2$, $\beta$, to maximize the expected achievable rate at Ru$_1$
while guaranteeing the QoS requirement at Ru$_2$.
For the MISO case, the expected achievable rate at Ru$_1$ is given by
\begin{align}\label{MISO-Ru1}
R_{_{\mathrm{Ru}_1}}\!(\bw_1,\beta)=\left\{\begin{array}{ll}
                       \bar{R}_{_{\mathrm{Ru}_1}}(\bw_1,\beta), & \mathrm{if}\ \tilde{g}_{0}>\tilde{g}_{1},\\
                       \frac{1}{2}\log_2\left(1+\rho_1\tilde{g}_{1}\right), & \mathrm{otherwise},
                     \end{array}\right.
\end{align}
where $\tilde{g}_{0}=\|\bh_0\|^2$, $\tilde{g}_{1}=\|\bh_1\|^2$ and $\bar{R}_{_{\mathrm{Ru}_1}}(\bw_1,\beta)$ is given by
\begin{align}\label{R1-MISO}
\!\bar{R}_{_{\mathrm{Ru}_1}}&\!(\bw_1,\beta)\!\!=\!\!\frac{\alpha}{2}\!\min\!\left[\!\log_2\!\left(\!1\!+\!\rho_1|\bh_{1}^\HH\bw_1|^2\!+\!\frac{\beta\rho_2 g_{21}}{(\!1\!-\!\beta\!)\rho_2 g_{21}\!+\!1\!}\!\right)\!,\right.\nonumber\\
&\left.\log_2\!\left(\!1\!+\!\rho_1|\bh_0^\HH\bw_1|^2\!\right)\vphantom{\log_2\!\left(1\!+\!\rho_1|\bh_{1}^\HH\bw_1|^2\!+\!\frac{\beta\rho_2 g_{21}}{(1\!-\!\beta)\rho_2 g_{21}\!+\!1}\right)}\right]\!\!+\!\!\frac{\!1\!\!-\!\!\alpha}{2}\!\log_2\!\left(\!1\!+\!\rho_1|\bh_{1}^\HH\bw_1|^2\right)\!.
\end{align}
The part related with both $\bw_1$ and $\beta$ in \eqref{R1-MISO} is defined by
\begin{align}\label{Q_t_MISO_1}
Q_{_{\mathrm{Tu}_1}}\!(\bw_1,\beta)\!=&\!\frac{1}{2}\min\!\left[\!\log_2\!\left(\!1\!\!+\!\rho_1|\bh_{1}^\HH\bw_1|^2\!\!+\!
\frac{\beta\rho_2g_{21}}{(\!1\!-\!\beta\!)\rho_2 g_{21}\!+\!1\!}\right)\right.,\nonumber\\
&\left.\log_2\!\left(1\!+\!\rho_1|\bh_0^\HH\bw_1|^2\right)\vphantom{\log_2\!\left(1\!+\!\rho_1|\bh_{1}^\HH\bw_1|^2\!+\!
\frac{\beta\rho_2g_{21}}{(1\!-\!\beta)\rho_2 g_{21}\!+\!1}\right)}\right].
\end{align}
Similar to SISO case, the expected achievable rate at Ru$_1$ is an increasing function of $P_1$ and hence, the maximum transmit power $P_1$ is always optimal for MISO case.

For the case that Tu$_2$ helps the transmission of Tu$_1$, if the receiving rate at Ru$_2$ in $t_1$, $R_{12}(\bw_1)$, is greater than $Q_{_{\mathrm{Tu}_1}}\!(\bw_1,\beta)$, i.e., $R_{12}(\bw_1)=\frac{1}{2}\log_2\big(1+\rho_1|\bh_{12}^\HH\bw_1|^2\big)\geq Q_{_{\mathrm{Tu}_1}}\!(\bw_1,\beta)$, Ru$_2$ can decode $x_1$ in $t_1$ and hence, Ru$_2$ can apply SIC to eliminate the effect of $x_1$ in its received signal in $t_2$. Therefore, the achievable rate at Ru$_2$ is given by
\begin{align}\label{MISO-Ru2}
\!\!\!\!\!R_{_{Ru_2}}\!(\beta)\!\!=\!\!\left\{\begin{array}{ll}
                       \!\!\!\!R_{_{\mathrm{Ru}_2}}^\mathrm{SIC}(\beta), ~~\mathrm{if}\ R_{12}(\!\bw_1)\!\!\geq\!\! Q_{_{\mathrm{Tu}_1}}\!(\!\bw_1\!,\!\beta),\tilde{g_0}\!\!>\!\!\tilde{g_1}, \\
                       \!\!\!\!R_{_{\mathrm{Ru}_2}}^\mathrm{NSIC}(\beta),~\mathrm{if}\ R_{12}(\!\bw_1\!)\!\!<\!\!Q_{_{\mathrm{Tu}_1}}\!(\bw_1\!,\!\beta),\tilde{g_0}\!\!>\!\!\tilde{g_1}, \\
                       \!\!\!\!\frac{1}{2}\!\log_2(\!1\!\!+\!\rho_2g_2\!),~\mathrm{otherwise},
                     \end{array}\right.\!\!
\end{align}
where $R_{_{\mathrm{Ru}_2}}^\mathrm{SIC}(\beta)$ and $R_{_{\mathrm{Ru}_2}}^\mathrm{NSIC}(\beta)$ are given in \eqref{R2-NSIC-SIC}.

For the MISO case, the optimal beamformer at Tu$_1$, $\bw_1$, and the power allocation for cooperation at Tu$_2$, $\beta$, that maximize the expected achievable rate at Ru$_1$ while guaranteeing QoS requirement at Ru$_2$ are obtained by solving the following joint optimization problem
\begin{subequations}\label{MISO-1}
\begin{align}
\mathbf{P2}~:~\max_{\bw_1, 0\leq \beta \leq 1} &~ R_{_{\mathrm{Ru}_1}}\!(\bw_1,\beta)\\
\mathrm{s.t.} & ~R_{_{\mathrm{Ru}_2}}\!(\beta)\geq Q,~~\bw_1^\HH\bw_1\leq 1.
\end{align}
\end{subequations}

We define the constant values for given channels, $v_1,\ v_2$ and $\phi_1$ as
\begin{align}
v_1\triangleq \|\bPi_{\mathbf{h}_0}\bh_1\|^2,\ v_2\triangleq \|\bPi_{\mathbf{h}_0}^\bot\bh_1\|^2,\ 
\phi_1\triangleq \rho_{1}\left(1+\frac{1}{\rho_2 g_{21}}\right).\nonumber
\end{align}
Then, the optimal structure of beamformer $\bw_1$ can be obtained by the following lemma.
\begin{lemma}\cite[Lemma 1]{ryu2015trust}\label{lemma_ryu}
The optimal beamformer at Tu$_1$ that maximizes the expected achievable rate at Ru$_1$ can be represented by
\begin{align}\label{w1-opt}
\bw_1^{\mathrm{opt}}=\sqrt{\eta}\bw_0+\sqrt{1-\eta}\bw_0^\bot,
\end{align}
where $\bw_0=\frac{\bPi_{\mathbf{h}_0}\bh_1}{\left\|\bPi_{\mathbf{h}_0}\bh_1\right\|}$, $\bw_0^\bot=\frac{\bPi_{\mathbf{h}_0}^\bot\bh_1}{\left\|\bPi_{\mathbf{h}_0}^\bot\bh_1\right\|}$ and $\eta$ is a constant in the range of $\frac{v_1}{v_1+v_2}\leq\eta\leq1$.
\end{lemma}

\begin{IEEEproof}
The proof of Lemma \ref{lemma_ryu} can be referred to \cite{ryu2015trust}.
\end{IEEEproof}

In Lemma \ref{lemma_ryu}, it is difficult to present $\eta$ that maximizes $\bar{R}_{_{\mathrm{Ru}_1}}(\bw_1,\beta)$ in closed form and hence, the optimal $\eta$ should be found by exhaustive search. However, from the numerically obtained beamformer $\bw_1$, we cannot get the insight on the effect of trust degree on $\bar{R}_{_{\mathrm{Ru}_1}}(\bw_1,\beta)$. Hence, in order to obtain $\bw_1$ in closed form, the approximated expected achievable rate at Ru$_1$ can be obtained by high signal-to-noise ratio (SNR) approximation\footnote{In order to obtain the closed-form beamformer, we adopt the high SNR approximation. However, it does not mean that we assume the high SNR configuration in our system model.}, as
\begin{align}\label{R1-MISO-app}
\tilde{R}_{_{\mathrm{Ru}_1}}&\!(\bw_1,\beta)\!\approx\frac{\alpha}{2}\min\!
\left[\!\log_2\!\left(\rho_1|\bh_{1}^\HH\bw_1|^2\!+\!\frac{\beta\rho_2 g_{21}}{(1\!-\!\beta\!)\rho_2 g_{21}\!+\!1}\right)\!,\right.\nonumber\\
&\left.\log_2\!\left(\!\rho_1|\bh_0^\HH\bw_1|^2\right)\vphantom{\log_2\!\left(\!\rho_1|\bh_{1}^\HH\bw_1|^2\!+\!\frac{\beta\rho_2 g_{21}}{(1\!-\!\beta)\rho_2 g_{21}\!+\!1}\right)}\right]\!\!+\!\frac{\!1\!-\!\alpha}{2}\log_2\!\left(\!\rho_1|\bh_{1}^\HH\bw_1|^2\right).
\end{align}

For given $\beta$, the transmit beamformer that maximizes the approximated expected achievable rate at Ru$_1$ can be obtained by the following theorem.
\begin{theorem}\label{theorem2}
For given trust degree $\alpha$ and the power allocation $\beta$, the transmit beamformer of Tu$_1$ that maximizes $\tilde{R}_{_{\mathrm{Ru}_1}}\!(\bw_1)$ is obtained by
\begin{align}\label{w_op}
\bw_1^\star=\sqrt{\eta^\star}\bw_0+\sqrt{1-\eta^\star}\bw_0^\bot,
\end{align}
where $\eta^\star$ is given by \eqref{eta}.
\begin{figure*}[!t]
\begin{align}\label{eta}
\eta^\star\!=\!\left\{\!\!\begin{array}{ll}
        \eta_1\!=\!\frac{v_1}{v_1+v_2},&\mathrm{if}\ \beta<\underline{\beta}=\frac{\{v_1\tilde{g}_{0}-(v_1+v_2)^2\}
        \phi_1}{v_1+v_2+\{v_1\tilde{g}_{0}\!-\!(v_1\!+\!v_2)^2\}\rho_1}, ~\tilde{g}_{0}\!>\!\frac{(v_1\!+\!v_2)^2}{v_1}, \\
        \eta_2\!=\!\frac{v_1+2v_2\alpha+\sqrt{v_1^2+4v_1v_2\alpha(1-\alpha)}}{2(v_1+v_2)},
        &\mathrm{if}~\left\{\begin{array}{ll}
        \beta>\overline{\beta}=\frac{(\tilde{g}_{0}-v_1)\phi_1}{1+(\tilde{g}_{0}-v_1)\rho_1}, ~\tilde{g}_{0}\!\geq\!v_1, \\
         \tilde{g}_{0}<v_1,
         \end{array}\right.\\
        \min\left\{\eta_2,~\eta_3(\beta)\right\},&\mathrm{otherwise},
      \end{array}\right.
\end{align}
where
\begin{align}
&\!\!\eta_{3}(\beta)\!=\!\frac{v_2(v_1\!+\!v_2\!+\!\tilde{g}_{0}\!)\!+\!m_1(\beta)(\tilde{g}_{0}\!\!-\!\!(v_1\!+\!v_2))
\!+\!2\sqrt{\!v_1\!v_2\{v_2\tilde{g}_{0}\!+\!m_1(\beta)(\tilde{g}_{0}\!-\!(v_1\!+\!v_2)\!-\!m_1(\beta))\}}}
{(\tilde{g}_{0}-v_1)^2+v_2(2v_1+v_2+2\tilde{g}_{0})},\\
&\!\!m_1(\beta)\!=\!\frac{\beta}{\phi_1-\beta\rho_1}.\label{eta3}
\end{align}
\hrulefill
\end{figure*}

\end{theorem}
\begin{IEEEproof}
The proof is presented in Appendix \ref{app2}.
\end{IEEEproof}

\begin{remark}
From Theorem \ref{theorem2}, we can see that the direction of beamformer at Tu$_1$, $\eta$, is affected by both the trust degree, $\alpha$ and the power allocation for cooperation, $\beta$.
First, when
the power allocation for cooperation is very small such as $\beta<\underline{\beta}$,
the expected achievable rate enhancement from the cooperation with Tu$_2$ is small even if $\alpha$ is large.
Hence, in this case, the beamformer at Tu$_1$ is designed to maximize the direct link from
Tu$_1$ to Ru$_1$ regardless of $\alpha$.
On the other hand, for high $\beta$ as $\beta > \overline{\beta}$,
since the expected achievable rate enhancement is large enough due to high $\beta$,
the direction of beamformer mainly depends on the trust degree, $\alpha$.
Hence, when Tu$_2$ helps the transmission of Tu$_1$ with high probability, equivalently $\alpha$ is high,
the direction of beamformer is steered toward $\mathbf{h}_{0}$ based on $\eta_2$, which is an increasing function with $\alpha$ to fully exploit the cooperation of Tu$_2$.
Otherwise, the direction of beamformer has to be properly steered based both $\alpha$ and $\beta$.
When $\alpha$ is relatively high compared to $\beta$ such as $\eta_2>\eta_3(\beta)$,
the direction of beamformer tends to be steered toward $\mathbf{h}_{0}$ rather than $\mathbf{h}_{1}$
and vice versa.
\end{remark}

In Theorem \ref{theorem2}, we show that beamformer $\bw_1$ can be represented according to $\beta$.
Thus, the joint optimization problem of $\bw_1$ and $\beta$ can be simplified by the optimization problem of a single parameter $\beta$, as
\begin{subequations}\label{MISO-11}
\begin{align}
\mathbf{P2-1}~:~\max_{0\leq \beta \leq 1} &~ R_{_{\mathrm{Ru}_1}}\!(\beta)\\
\mathrm{s.t.} & ~R_{_{\mathrm{Ru}_2}}\!(\beta)\geq Q.
\end{align}
\end{subequations}

For the general case, it is hard to directly obtain the optimal $\beta^\star$ from $\mathbf{P2-1}$ because
$\mathbf{P2-1}$ is non-convex with respect to $\beta$.
Thus, in the following corollaries, we obtain the optimal power allocation for some special cases.
In the following corollaries, we assume $\rho_1=\rho_2$ for simplicity.

\begin{corollary}\label{coro-eta-beta}
When the channel from Tu$_1$ to Tu$_2$ is very strong such as
$\tilde{g}_{0}\geq \overline{g_{0}}\triangleq\frac{\tilde{g}_{1}\left(\tilde{g}_{1}+g_{21}\right)}{v_1}$,
the optimal power allocation for cooperation that maximizes the approximated achievable rate at Ru$_1$ can be represented by
\begin{align}
  \beta^\star=\left\{\begin{array}{ll}
               \min\big\{\tilde{\beta}_{2},\beta_{Q_1}\big\}, & \mathrm{if}\ v_3\geq\tilde{g}_{1}^{2},\\
               \beta_{Q_2}, & \mathrm{otherwise},
             \end{array}\right.
\end{align}
where $\tilde{\beta}_{2}$ and $v_3$ are given by
\begin{align}
\tilde{\beta}_{2}=\frac{\left(v_3-\tilde{g}_{1}^{2}\right)\phi_1}
{\tilde{g}_{1}+\left(v_3-\tilde{g}_{1}^{2}\right)\rho_1},~~
v_3=\big|\bh_{12}^\HH\bh_1\big|^{2},
\end{align}
and $\beta_{Q_1}$ and $\beta_{Q_2}$ are given in \eqref{betaQ1} and \eqref{betaQ2}, respectively.
\end{corollary}

\begin{IEEEproof}
For $\tilde{g}_{0}\geq \overline{g_{0}}$, since we have $\underline{\beta}\geq1$, for all feasible $\beta$
in $0\leq \beta \leq 1$, the optimal beamformer is given by $\bw_1^\star(\eta_{1})=\sqrt{\eta_1}\bw_0+\sqrt{1-\eta_1}\bw_0^\bot=\frac{\mathbf{h}_{1}}{\sqrt{\tilde{g}_{1}}}$ and
the approximated achievable rate at Ru$_1$ is represented by \eqref{R1-g}
with $\eta_1$.
For this case, the condition to apply SIC at Ru$_2$ is
\begin{align}
\!\!\!\!\!\!R_{12}(\bw_1^\star)&=\frac{1}{2}\log_{2}\left(1+\frac{\rho_{1}v_{3}}{\tilde{g}_{1}}\right)
\nonumber\\
&\geq\frac{1}{2}\log_2\left(1+\rho_1\big(\tilde{g}_{1}+m_1(\beta)\big)\right)\nonumber\\
&={Q}_{\mathrm{Tu}_{1}}(\bw_1^\star,\beta)~
\!\!\Rightarrow\!\!\beta\leq \tilde{\beta}_{2}.
\end{align}
If $v_3<\tilde{g}_{1}^{2}$, Ru$_2$ cannot apply SIC for all $\beta$ and hence, for $v_3\geq\tilde{g}_{1}^{2}$,
the condition to apply SIC at Ru$_2$ while guaranteeing QoS requirement is obtained by
$\beta\leq \min\big\{\tilde{\beta}_{2},\beta_{Q_1}\big\}$.
Since \eqref{R1-g} is an increasing function of $\beta$, the optimal power allocation is obtained by
$\beta^{\star}= \min\big\{\tilde{\beta}_{2},\beta_{Q_1}\big\}$.

Otherwise, Ru$_2$ cannot apply SIC for all $\beta$ and hence, the power allocation to guarantee QoS without SIC is obtained by $\beta^{\star}=\beta_{Q_2}$.
\end{IEEEproof}

\begin{corollary}\label{coro-eta-beta2}
When the channel from Tu$_1$ to Tu$_2$ is very weak such as
$\tilde{g}_{0}<{v_1}$,
the optimal power allocation for cooperation that maximizes the approximated achievable rate at Ru$_1$ can be represented by
\begin{align}
  \beta^\star=\left\{\begin{array}{ll}
               \beta_{Q_1}, & \mathrm{if}\ v_4\geq v_5,\\
               \beta_{Q_2}, & \mathrm{otherwise},
             \end{array}\right.
\end{align}
where $v_4$ and $v_5$ are given by
$v_4=\big|\bh_{12}^\HH\bw_1^{\star}(\eta_{2})\big|^{2}$ and $v_5=\big|\bh_{0}^\HH\bw_1^{\star}(\eta_{2})\big|^{2}$.
\end{corollary}
\begin{IEEEproof}
If $\tilde{g}_{0}< v_1$, for all feasible $\beta$
in $0\leq \beta \leq 1$, the optimal beamformer is given by $\bw_1^\star(\eta_{2})=\sqrt{\eta_2}\bw_0+\sqrt{1-\eta_2}\bw_0^\bot$.
Hence, the remaining part of the proof of Corollary \ref{coro-eta-beta2} can be obtained in the similar way to the proof of Corollary \ref{coro-eta-beta}.
\end{IEEEproof}

From Lemma \ref{lemma_ryu}, Theorem \ref{theorem2} and high SNR approximation, the joint optimization problem to design beamformer at Tu$_1$, $\bw_1$, and the power allocation at Tu$_2$, $\beta$, is simplified into the optimization problem with a single parameter $\beta$.
For some special case, the optimal $\beta$ can be obtained in a closed form.
For general case, the optimal power allocation for cooperation can be obtained by one dimensional search from 0 to 1, which is much simpler than solving the joint optimization problem.

\subsection{SIMO case ($N_1=1$ and $N_2\geq 2$)}
In this subsection, we consider the SIMO case, where Tu$_1$ equips with a single antenna (i.e., $N_1=1$) but Tu$_2$ is equipped with $N_2\geq2$ antennas and hence, the SIMO channel is formed between Tu$_1$ and Tu$_2$.
For the SIMO case, we jointly design the transmit beamformers at Tu$_2$, $\bw_{21}$ and $\bw_{22}$, to maximize the expected achievable rate at Ru$_1$. For this case, the expected achievable rate at Ru$_1$ in SIMO case is given by
\begin{eqnarray}
R_{_{\mathrm{Ru}_1}}(\bw_{21},\bw_{22})=\left\{\begin{array}{ll}
                       \bar{R}_{_{\mathrm{Ru}_1}}(\bw_{21},\bw_{22}), & \text{if}\ \tilde{g}_0>g_1,\\
                       \frac{1}{2}\log_2\left(1+\rho_1g_1\right), & \text{otherwise},
                     \end{array}\right.
\end{eqnarray}
where $\tilde{g}_0>g_1$ means the channel condition between Tu$_1$ and Tu$_2$ is better than the direct channel from Tu$_1$ to Ru$_1$, thus the cooperation of Tu$_2$ can improve the achievable rate at Ru$_1$, and $\bar{R}_{_{\mathrm{Ru}_1}}(\bw_{21},\bw_{22})$ is given by
\begin{align}
\!\!\!\!\bar{R}_{_{\mathrm{Ru}_1}}\!(\bw_{21}\!,\!\bw_{22}\!)&\!\!=\!\!
\frac{\alpha}{2}\text{min}\!\left[\!\log_2\!\!\left(\!\!1\!\!+\!\rho_1g_1
\!+\!\frac{|\bh_{21}^\HH\bw_{21}|^2}{|\bh_{21}^\HH\bw_{22}|^2\!+\!\sigma^2}\!\right),\right.\nonumber\\
&\left.\log_2\!\left(\!1\!+\!\rho_1\tilde{g}_0\!\right)\!\vphantom{\!\log_2\!\!\left(\!\!1\!\!+\!\rho_1g_1
\!+\!\frac{|\bh_{21}^\HH\bw_{21}|^2}{|\bh_{21}^\HH\bw_{22}|^2\!+\!\sigma^2}\!\right)}\right]\!\!+\!\!\frac{\!1\!-\!\alpha\!}{2}\log_2\left(\!1\!+\!\rho_1g_1\!\right).\label{SIMO-Ru1}
\end{align}
We define the first term in \eqref{SIMO-Ru1} as
\begin{align}\label{Q_t_SIMO}
Q_{_{\mathrm{Tu}_1}}\!(\bw_{21},\bw_{22})\!=&
\frac{1}{2}\min\!\left[\log_2\!\left(\!1\!+\!\rho_1g_1
\!+\!\frac{|\bh_{21}^\HH\bw_{21}|^2}{|\bh_{21}^\HH\bw_{22}|^2\!+\!\sigma^2}\right),\right.\nonumber\\
&\left.\log_2\!\left(1\!+\!\rho_1\tilde{g}_0\right)\!\vphantom{\log_2\!\left(\!1\!+\!\rho_1g_1
\!+\!\frac{|\bh_{21}^\HH\bw_{21}|^2}{|\bh_{21}^\HH\bw_{22}|^2\!+\!\sigma^2}\right)}\right].
\end{align}

For the case that Tu$_2$ helps the transmission of Tu$_1$, if the rate achieved at Ru$_2$ in $t_1$, $R_{12}$, is greater than $Q_{_{\mathrm{Tu}_1}}(\bw_{21},\bw_{22})$, i.e., $R_{12}=\frac{1}{2}\log_2\left(1+\rho_1g_{12}\right)\geq Q_{_{\mathrm{Tu}_1}}(\bw_{21},\bw_{22})$, Ru$_2$ can decode $x_1$ in $t_1$ and thus, Ru$_2$ can employ SIC to eliminate the effect of $x_1$ in its received signal in $t_2$. Therefore, the achievable rate at Ru$_2$ is given by
\begin{align}\label{Ru_2}
\!R_{_{\mathrm{Ru}_2}}\!(\!\bw_{21}\!,\!\bw_{22}\!)\!\!=\!\!\left\{\!\!\begin{array}{ll}
                       \!\!R_{_{\mathrm{Ru}_2}}^{\text{SIC}}(\!\bw_{22}\!), ~\text{if}\left\{\begin{array}{ll}
                        \!\!\!R_{12}\!\!\geq\!\!Q_{_{\mathrm{Tu}_1}}\!(\!\bw_{21}\!,\!\bw_{22}\!)\\ \!\!\!\tilde{g}_0\!>\!g_1
                        \end{array}\right.,\!\!\! \\
                       \!\!R_{_{\mathrm{Ru}_2}}^{\text{NSIC}}(\!\bw_{21}\!,\!\bw_{22}\!), ~\text{if}\left\{\begin{array}{ll}
                       \!\!\!R_{12}\!\!<\!\!Q_{_{\mathrm{Tu}_1}}\!\!(\!\bw_{21}\!,\!\bw_{22}\!)\\
                       \!\!\!\tilde{g}_0\!>\!g_1
                       \!\! \end{array}\right.\!\!\!\!\!,\!\!\! \\
                       \!\!\frac{1}{2}\log_2(\!1\!+\!\rho_2\tilde{g}_2\!),~\text{otherwise},
                     \end{array}\right.\!\!\!\!\!\!\!
\end{align}
where $R_{_{\mathrm{Ru}_2}}^{\text{SIC}}(\bw_{22})$ and $R_{_{\mathrm{Ru}_2}}^{\text{NSIC}}(\bw_{21},\bw_{22})$ are respectively given by
\begin{align}
\begin{array}{ll}
R_{_{\mathrm{Ru}_2}}^{\text{SIC}}(\bw_{22})=\frac{1}{2}\log_2\left(1+\frac{|\bh_2^\HH\bw_{22}|^2}{\sigma^2}\right),\\
R_{_{\mathrm{Ru}_2}}^{\text{NSIC}}(\bw_{21},\bw_{22})=\frac{1}{2}\log_2\left(1+\frac{|\bh_2^\HH\bw_{22}|^2}{|\bh_2^\HH\bw_{21}|^2+\sigma^2}\right).\label{R2-NSIC-SIC-MISO}
\end{array}
\end{align}

For the SIMO case, to maximize the expected achievable rate at Ru$_1$ while guaranteeing QoS of Ru$_2$, the beamformers at Tu$_2$, $\bw_{21}$ and $\bw_{22}$,
are jointly optimized by the following problem
\begin{subequations}\label{OP3}
\begin{align}
\mathbf{P3}~:~\max_{\bw_{21},\bw_{22}} &~ R_{_{\mathrm{Ru}_1}}\!(\bw_{21},\bw_{22})\\
\mathrm{s.t.} & ~R_{_{\mathrm{Ru}_2}}(\bw_{21},\bw_{22})\geq Q,&\\
&~ \bw_{21}^\HH\bw_{21}+\bw_{22}^\HH\bw_{22}\leq P_2,
\end{align}
\end{subequations}
where the power allocation at Tu$_2$ is embedded in the beamformer design.

By considering the condition whether Ru$_2$ applies SIC, $\mathbf{P3}$ in \eqref{OP3} can be divided by two subproblems. For the case that Ru$_2$ applies SIC, the optimization problem $\mathbf{P3-1}$ is given by
\begin{subequations}\label{P3-1}
\begin{align}
\!\mathbf{P3-1}:\!\!\!\max_{\bw_{21},\bw_{22}} ~&R_{_{\mathrm{Ru}_1}}\!(\bw_{21},\bw_{22})&\label{P3-1-OBJ}\\
\!\mathrm{s.t.}
~&\!\frac{1}{2}\!\log_2\left(\!1\!\!+\!\!\rho_1g_{12}\right)\!\geq\! Q_{_{\mathrm{Tu}_1}}\!(\!\bw_{21}\!,\!\bw_{22}\!),&\label{P3-1-C1}\\
&\frac{1}{2}\log_2\left(1\!+\!\frac{|\bh_2^\HH\bw_{22}|^2}{\sigma^2}\right)\geq Q,&\label{P3-1-C2}\\
~&\bw_{21}^\HH\bw_{21}+\bw_{22}^\HH\bw_{22}\leq P_2,&\label{P3-1-C3}
\end{align}
\end{subequations}
where constraint \eqref{P3-1-C1} is the condition that Ru$_2$ decodes the data from Tu$_1$ in $t_1$ and applies SIC to cancel it in $t_2$. Similarly, the case that Ru$_2$ does not apply SIC, $\mathbf{P3-2}$, is given by
\begin{subequations}\label{P3-2}
\begin{align}
\!\mathbf{P3-2}:\!\!\!\max_{\bw_{21},\bw_{22}} ~&R_{_{\mathrm{Ru}_1}}\!(\bw_{21},\bw_{22})&\label{P3-2-OBJ}\\
\!\mathrm{s.t.}
~&\!\frac{1}{2}\!\log_2\!\left(\!1\!+\!\rho_1g_{12}\right)\!<\! Q_{_{\mathrm{Tu}_1}}\!(\!\bw_{21}\!,\!\bw_{22}\!),&\label{P3-2-C1}\\
&\!\frac{1}{2}\!\log_2\!\left(\!1\!+\!\frac{\!|\bh_2^\HH\bw_{22}|^2}{|\bh_2^\HH\bw_{21}|^2+
\sigma^2}\right)\!\geq\!Q,&\label{P3-2-C2}\\
~&\bw_{21}^\HH\bw_{21}+\bw_{22}^\HH\bw_{22}\leq P_2.&\label{P3-2-C3}
\end{align}
\end{subequations}

For the relaying case, i.e., $\tilde{g}_0>g_1$, from \eqref{SIMO-Ru1}, $R_{_{\mathrm{Ru}_1}}\!(\bw_{21},\bw_{22})$ can be presented by two different forms according to $\bw_{21}$ and $\bw_{22}$ due to DF relaying constraint. Therefore, both $\mathbf{P3-1}$ and $\mathbf{P3-2}$ can be further divided into two subproblems with respect to $R_{_{\mathrm{Ru}_1}}\!(\bw_{21},\bw_{22})$. Hence, we can obtain the solution of $\mathbf{P3}$ in \eqref{OP3} by choosing the best solution from the solutions of four subproblems. Since all subproblems can be solved in a similar way, here we focus on one of the four subproblems. In the following, we consider the subproblem $\mathbf{P3-21}$, where Ru$_2$ does not apply SIC in $t_2$ and $Q_{_{\mathrm{Tu}_1}}(\bw_{21},\bw_{22})$ is
determined by
\begin{align}\label{QTu1_min_SIMO}
Q_{_{\mathrm{Tu}_1}}(\!\bw_{21}\!,\!\bw_{22}\!)\!=\!\frac{1}{2}\!\log_2\!\left(1\!+\!\rho_1g_1
\!+\!\frac{|\bh_{21}^\HH\bw_{21}|^2}{|\bh_{21}^\HH\bw_{22}|^2\!+\!\sigma^2}\right).
\end{align}
Therefore, for this case, the subproblem $\mathbf{P3-21}$ is represented by
\begin{subequations}\label{SOP_SIMO}
\begin{align}
\mathbf{\!P3\!\!-\!\!21\!}\!:\!\!\!\!\max_{\!\bw_{21}\!,\!\bw_{22}\!} ~&\!\!\!\!\!\alpha\!\log_2\!\!\left(\!\!\!1\!\!+\!\!\rho_1g_1
\!\!+\!\!\frac{\!|\bh_{21}^\HH\!\bw_{21}|^2}{\!|\bh_{21}^\HH\!\!\bw_{22}|^2\!\!+\!\!\sigma^2}\!\!\!\right)
\!\!\!+\!\!(\!1\!\!-\!\!\alpha\!)\!\log_2\!\!\left(\!1\!\!+\!\!\rho_1g_1\!\right)\!\!\!\!\!&\label{SOP-OF-SIMO}\\
\mathrm{\!s.t.}
\!\log_2&\!\!\left(\!1\!\!+\!\!\rho_1g_{12}\!\right)\!\!<\!\! \log_2\!\!\left(\!\!1\!\!+\!\!\rho_1g_1\!\!+\!
\!\frac{|\bh_{21}^\HH\bw_{21}|^2}{|\bh_{21}^\HH\!\!\bw_{22}\!|^2\!\!+\!\!\sigma^2\!}\!\right)\!\!,\!\!\!\!\!\!\!\!&\label{SOP-C1-SIMO}\\
\!\log_2&\!\!\left(\!1\!\!+\!\!\rho_1\tilde{g}_0\!\right)\!\!\geq\!\! \log_2\!\!\left(\!\!\!1\!\!+\!\!\rho_1g_1\!\!+\!\!
\!\frac{|\bh_{21}^\HH\bw_{21}|^2}{|\!\bh_{21}^\HH\!\!\bw_{22}\!|^2\!\!+\!\!\sigma^2\!}\!\!\right)\!\!,\!\!\!&\label{SOP-C2-SIMO}\\
\!\frac{1}{2}\!\!\log_2\!\!&\left(\!\!1\!+\!\frac{|\bh_2^\HH\bw_{22}|^2}{|\bh_2^\HH\bw_{21}|^2+
\sigma^2}\right)\geq Q,&\label{SOP-C3-SIMO}\\
\!\!\!\!\bw_{21}^\HH~&\!\!\!\bw_{21}\!+\!\bw_{22}^\HH\bw_{22}\leq P_2,&\label{SOP-C4-SIMO}
\end{align}
\end{subequations}
where the constraint \eqref{SOP-C1-SIMO} is the condition that Ru$_2$ cannot apply SIC in $t_2$ and the constraint \eqref{SOP-C2-SIMO} is the condition to satisfy \eqref{QTu1_min_SIMO} referring to \eqref{Q_t_SIMO}. Notice that constant $\frac{1}{2}$ is ignored in \eqref{SOP-OF-SIMO} without changing the property of the problem. From the observation of $\mathbf{P3-21}$, the second term of the objective function, i.e., $(1\!-\!\alpha)\log_2\!\left(1\!+\!\rho_1g_1\!\right)$, is a constant, which can be ignored to obtain the solution. Then, we can see that for given $\alpha$, the right hand sides (RHSs) of constraints \eqref{SOP-C1-SIMO} and \eqref{SOP-C2-SIMO} are equal to the objective function of \eqref{SOP-OF-SIMO} to be maximized, i.e., $\log_2\!\left(1\!+\!\rho_1g_1 \!+\!\frac{|\bh_{21}^\HH\bw_{21}|^2}{|\bh_{21}^\HH\bw_{22}|^2\!+\!\sigma^2}\right)$.
Hence, we note that if the problem $\mathbf{P3-21}$ is feasible, the constraint \eqref{SOP-C1-SIMO} is always hold. Otherwise, $\mathbf{P3-21}$ is infeasible and the solution is obtained by the other subproblems.
In addition, if the constraint \eqref{SOP-C2-SIMO} is not hold, i.e.,
$\log_2\left(1\!+\!\rho_1\tilde{g}_0\right)< \log_2\left(1\!+\!\rho_1g_1\!+
\!\frac{|\bh_{21}^\HH\bw_{21}|^2}{|\bh_{21}^\HH\bw_{22}|^2+\sigma^2}\right)$, from \eqref{SIMO-Ru1},
we observe that the expected achievable rate is bounded by constant, which is independent from
$\bw_{21}$ and $\bw_{22}$. In other word, Ru$_1$ can achieve at least constant rate in \eqref{SIMO-Ru1}
if the constraint \eqref{SOP-C2-SIMO} is not hold.
Hence, we can solve $\mathbf{P3-21}$ without considering the constraint \eqref{SOP-C2-SIMO} and after solving $\mathbf{P3-21}$ without \eqref{SOP-C2-SIMO}, we can check whether the constraint \eqref{SOP-C2-SIMO} is hold or not for the obtained solution.
If \eqref{SOP-C2-SIMO} is hold, the expected achievable rate at Ru$_1$ is determined by the obtained solution and otherwise, the expected achievable rate is determined by the constant rate in \eqref{SIMO-Ru1}. Therefore, by removing constraints \eqref{SOP-C1-SIMO} and \eqref{SOP-C2-SIMO}, the subproblem $\mathbf{P3-21}$ can be equivalently rewritten as
\begin{subequations}\label{SOP1-SIMO}
\begin{align}
\!\!\!\!\mathbf{\!P3\!\!-\!\!21'\!}:\max_{\bw_{21},\bw_{22}} ~&\frac{|\bh_{21}^\HH\!\bw_{21}|^2}{|\bh_{21}^\HH\!\bw_{22}|^2\!+\!\sigma^2}&\label{SOP1-OF-SIMO}\\
\!\!\mathrm{s.t.}\!
~&|\bh_2^\HH\!\bw_{22}|^2\!\!\geq\!\! (\!4^Q\!\!-\!\!1)\!(|\bh_2^\HH\!\bw_{21}|^2\!\!+\!\!\sigma^2),\!\!\!&\label{SOP1-C1-SIMO}\\
~&\bw_{21}^\HH\bw_{21}+\bw_{22}^\HH\bw_{22}\leq P_2.&\label{SOP1-C2-SIMO}
\end{align}
\end{subequations}
Since the problem $\mathbf{P3-21'}$ is non-convex and $\bw_{21}$, $\bw_{22}$ are still coupled in the constraints, it is hard to directly obtain the solution of $\mathbf{P3-21'}$ in its current form.
To solve $\mathbf{P3-21'}$ by decoupling $\bw_{21}$ and $\bw_{22}$, we apply block coordinate update (BCU) method to update $\bw_{21}$ (or $\bw_{22}$) while fixing $\bw_{22}$ (or $\bw_{21}$) at one iteration, and optimize $\bw_{22}$ (or $\bw_{21}$) based on the newly updated $\bw_{21}$ (or $\bw_{22}$) at the next iteration.
Thus, the expected achievable rate at Ru$_1$ is maximized by optimizing $\bw_{21}$ and $\bw_{22}$ iteratively. Employing the semidefinite relaxation (SDR) technique \cite{Luo2010SPM} and giving $\bw_{22}^k$ at $k+1$-th iteration, the relaxation of \eqref{SOP1-SIMO} can be rewritten as
\begin{subequations}\label{SOP1-SIMO-1}
\begin{align}
\mathbf{\!P3\!\!-\!\!21'\!\!-\!\!1\!}\!:\max_{\bW_{21}\succeq\bzero} ~&\text{tr}(\bW_{21}\bh_{21}\bh_{21}^\HH)&\label{SOP1-OF-SIMO-1}\!\!\\
\mathrm{s.t.}
~&\!\!|\bh_2^\HH\!\bw_{22}^k\!|^2\!\!\geq\!\! (\!4^Q\!\!\!-\!\!1\!)\!\!\left[\!\text{tr}(\bW_{21}\!\bh_2\bh_2^\HH)\!\!+\!\!\sigma^2\!\right]\!\!,\!\!\!\!\!\!\!\!\!&\label{SOP1-C1-SIMO-1}\\
~&\text{tr}(\bW_{21})+(\bw_{22}^{k})^\HH\bw_{22}^k\leq P_2,&\label{SOP1-C2-SIMO-1}
\end{align}
\end{subequations}
where we discard the constraint $\text{rank}(\bW_{21})=1$. The relaxed problem \eqref{SOP1-SIMO-1} can be solved conveniently by existing solvers, such as CVX \cite{grant2011cvx}. It is noted that the sufficient and necessary condition for the equivalence of problems $\mathbf{P3\!-\!21'\!-\!1}$ and $\mathbf{P3-21'}$ with given $\bw_{22}^k$, is that the optimal $\bW_{21}^{*,k+1}$ obtained at $k+1$-th iteration of $\mathbf{P3\!-\!21'\!-\!1}$ is rank-one, i.e, $\bW_{21}^{*,k+1}=\bw_{21}^{*,k+1}(\bw_{21}^{*,k+1})^\HH$, which can be guaranteed by the following lemma.

\begin{lemma}\label{lemma_zhang}
\cite[Theorem 2.2]{AiHZ11} Let $\bA_i \in \mathbb{C}^{n\times n}, i\in \mathcal{I}=\{1,2,3\}$, be a Hermitian matrix and $\bX\in \mathcal{H}^n_+$ be a nonzero Hermitian positive semidefinite matrix. If $\text{rank}(\bX)\geq 2$, we can find a rank-one matrix $\bx\bx^\HH$ in polynomial-time such that $\text{tr}(\bA_i\bx\bx^\HH)=\text{tr}(\bA_i\bX),i\in\mathcal{I}$.
\end{lemma}
\begin{IEEEproof}
  The proof of Lemma \ref{lemma_zhang} can be referred to \cite{AiHZ11}.
\end{IEEEproof}

After achieving the rank-one solution $\bw_{21}^{*,k+1}$ from $\bW_{21}^{*,k+1}$ resorting to Lemma \ref{lemma_zhang} if $\mathbf{P3-21'}$ is feasible, we plug it into $\mathbf{P3-21'}$. In a similar way to $\mathbf{P3\!-\!21'\!-\!1}$, for given $\bw_{21}^{*,k+1}$, we solve the following problem
\begin{subequations}\label{SOP1-SIMO-2}
\begin{align}
\mathbf{\!P3\!\!-\!\!21'\!\!-\!\!2}\!:\!\!\!\!\!\min_{\bW_{22}\succeq\bzero} &\text{tr}(\bW_{22}\bh_{21}\bh_{21}^\HH)\\
\mathrm{s.t.}
\text{tr}(&\bW_{22}\bh_2\bh_2^\HH)\!\!\geq \!\! (\!4^Q\!\!\!-\!\!\!1)\!(|\bh_2^\HH\!\bw_{21}^{*,k\!+\!1}|^2\!\!+\!\!\sigma^2)\!,\!\!\!\!\!\label{SOP1-C1-SIMO-2}\\
&\!\!\!(\bw_{21}^{*,k+1})^\HH\bw_{21}^{*,k+1}\!\!+\!\text{tr}(\bW_{22})\!\!\leq\!\! P_2.\!\!\!\!\!\label{SOP1-C2-SIMO-2}
\end{align}
\end{subequations}

Similarly, we can obtain the rank-one $\bw_{22}^{*,k+1}$ from $\bW_{22}^{*,k+1}$ at $k+1$-th iteration based on Lemma \ref{lemma_zhang} if problem $\mathbf{P3-21'-2}$ is feasible. Consequently, we can iteratively obtain the optimal solution $(\bw_{21}^*,\bw_{22}^*)$ of $\mathbf{P3-21'}$. Then, by using $(\bw_{21}^*,\bw_{22}^*)$, we can check the feasibility of the constraints \eqref{SOP-C1-SIMO} and \eqref{SOP-C2-SIMO} of $\mathbf{P3-21}$, which are not considered to obtain $(\bw_{21}^*,\bw_{22}^*)$. If the constraints are feasible, the obtained solution $(\bw_{21}^*,\bw_{22}^*)$ can be a candidate of the optimal beamformers for SIMO case,
which are chosen among the solutions of four subproblems.
Otherwise, the optimal solution is obtained by solving the other subproblems.
The proposed iterative algorithm to solve $\mathbf{P3-21}$ is summarized in the TABLE \Rmnum{1}.

\begin{table}
\caption{The Proposed Algorithm For SIMO}
\begin{center}
\begin{tabular}{|l|}
\hline
\textbf{Initialization:} Generate feasible block variables $(\!\bw_{21}^0\!,\!\bw_{22}^0\!)$.\!\! \\
\text{S1}: \textbf{For} $k=1,\cdots,K$, where $K$ is the maximal iteration\\
\qquad\quad~ times, do S2--S3 until converge.\\
\textbf{Block Coordinate Update:}\\
\text{S2}: (1) Solve problem $\mathbf{P3\!-\!21'\!-\!1}$ with given $\bw_{22}^{*,k}$\\
\qquad\quad \textbf{If} the problem is feasible, do \text{S4} to obtain\\
\qquad\quad~~  the optimal $\bw_{21}^{*,k+1}$;\\
\qquad\quad \textbf{Else} $\bw_{21}^{*,k+1}:=\bw_{21}^{*,k}$.\\
\quad\ \ (2) Solve problem $\mathbf{\!P3\!-\!21'\!-\!2'\!}$ with given $\bw_{21}^{*,k+1}$\\
\qquad\quad \textbf{If} the problem is feasible, do \text{S4} to obtain\\
\qquad\quad~~ the optimal $\bw_{22}^{*,k+1}$;\\
\qquad\quad \textbf{Else} $\bw_{22}^{*,k+1}:=\bw_{22}^{*,k}$.\\
\textbf{Stopping Criteria:} Set $s^k=\frac{|\bh_{21}^\HH\bw_{21}^{*,k}|^2}{|\bh_{21}^\HH\bw_{22}^{*,k}|^2+\sigma^2}$.\\
\text{S3}:\ \textbf{If} $ \frac{|s^{k+1}-s^{k}|}{|s^{k+1}|}\leq \epsilon $, stop and return \\
\qquad~ $(\bw_{21}^{*,k+1}, \bw_{22}^{*,k+1}, s^{k+1})$, then do \text{S5};\\
\quad\ \ \textbf{Else} set $k:=k+1$ and go to \text{S2}.\\
\text{S4}: Set $\bX:=\bW_{21}^{*,k+1}$ (or $\bW_{22}^{*,k+1}$)\\
\quad\ \ \textbf{If} rank($\bX$)=1, the optimal solution is $\bw_{21}^{*,k+1}=\bx$\\
\qquad~  or $\bw_{22}^{*,k+1}=\bx$, where $\bX=\bx\bx^\HH$.\\
\quad\ \ \textbf{Else} employ Lemma \ref{lemma_zhang} to find a rank-one matrix $\bz\bz^\HH$,\\
\qquad~ the optimal solution is $\bw_{21}^{*,k+1}\!\!=\!\!\bz$ or $\bw_{22}^{*,k+1}\!\!=\!\!\bz$.\\
\textbf{Feasibility Check:} \\
\text{S5}: \textbf{If} \eqref{SOP-C1-SIMO} and \eqref{SOP-C2-SIMO} are feasible, $(\bw_{21}^{*,k+1}, \bw_{22}^{*,k+1})$\\
\qquad~ is a candidate solution;\\
\quad\ \ \textbf{Else} Solve the other subproblems.\\
  \hline
\end{tabular}
\end{center}
\end{table}

\subsection{MIMO case ($N_1\geq 2$ and $N_2\geq 2$)}
In this subsection, we consider that both Tu$_1$ and Tu$_2$ are equipped with $N_1\geq2$ and $N_2\geq2$ antennas and hence, the MIMO channel is formed between Tu$_1$ and Tu$_2$.
For the MIMO case, we jointly design the transmit beamformer at Tu$_1$, $\bw_1$, and
beamformers at Tu$_2$, $\bw_{21}$ and $\bw_{22}$, to maximize the expected achievable rate at Ru$_1$.
In this case, the relaying transmission of Tu$_2$ cannot improve the achievable rate
if the channel quality between Tu$_1$ and Tu$_2$ is worse than the direct channel from Tu$_1$ to Ru$_1$ such as
\begin{align}
\underset{\bw_{1}}{\max}~\log_{2}\left(\!1\!+\!\frac{\bw_{1}^\HH\bH_0^\HH\bH_0\bw_{1}}{\sigma^2}\right)\!\!=\!&
\log_{2}\left(1\!+\!\rho_1\lambda_{\max}(\bH_0^\HH\bH_0)\right)\nonumber
\\
\leq&\log_{2}\big(1+\rho_1\tilde{g}_1\big),
\end{align}
where $\lambda_{\max}(\mathbf{X})$ is the largest eigenvalue of $\mathbf{X}$. Therefore, for the MIMO case, the expected achievable rate at Ru$_1$ is given by
\begin{eqnarray}
R_{_{\mathrm{Ru}_1}}\!\!(\!\bw_1\!,\!\bw_{21}\!,\!\bw_{22}\!)\!\!=\!\!\left\{\begin{array}{ll}
                       \!\!\!\!\bar{R}_{_{\mathrm{Ru}_1}}\!\!(\!\bw_1\!,\!\bw_{21}\!,\!\bw_{22}\!)\!,\! \text{if}~\! \lambda_{\max}(\bH_0^\HH\!\bH_0\!)\!\!>\!\!\tilde{g}_1\!,\\
                       \!\!\!\!\frac{1}{2}\!\log_2\left(1\!\!+\!\!\rho_1\tilde{g}_1\right), \text{otherwise},
                     \end{array}\right.\!\!\!\!\!\!
\end{eqnarray}
where $\bar{R}_{_{\mathrm{Ru}_1}}(\bw_1,\bw_{21},\bw_{22})$ is given by
\begin{align}
\bar{R}_{_{\mathrm{Ru}_1}}\!&(\!\bw_1\!,\!\bw_{21}\!,\!\bw_{22}\!)\!\!=\!\!
\frac{\alpha}{2}\!\text{min}\!\left[\!\log_2\!\left(\!1\!\!+\!\!\frac{|\bh_{1}^\HH\bw_1|^2}{\sigma^2}
\!\!+\!\!\frac{|\bh_{21}^\HH\bw_{21}|^2}{|\bh_{21}^\HH\bw_{22}|^2\!\!+\!\!\sigma^2}\!\right)\!,\right.\nonumber\\
&\left.\log_2\!\left(\!1\!\!+\!\!\frac{\|\bH_0\bw_1\|^2}{\sigma^2}\!\right)\!\!\vphantom{\log_2\!\left(1\!+\!\frac{|\bh_{1}^\HH\bw_1|^2}{\sigma^2}
\!\!+\!\!\frac{|\bh_{21}^\HH\bw_{21}|^2}{|\bh_{21}^\HH\bw_{22}|^2\!+\!\sigma^2}\right)}\right]\!\!+\!\!\frac{1\!\!-\!\!\alpha}{2}\log_2\!\left(\!1\!\!+\!\!\frac{|\bh_{1}^\HH\bw_1|^2}{\sigma^2}\!\right).\!\!\!\label{Ru_1}
\end{align}
We define the first term in \eqref{Ru_1} as
\begin{align}\label{Q_t_MIMO}
Q_{_{\mathrm{Tu}_1}}\!\!(\!\bw_1\!,\!\bw_{21}\!,\!\bw_{22})\!\!=&\!\frac{1}{2}\!\!
\min\!\!\left[\!\log_2\!\left(\!1\!\!+\!\!\frac{|\bh_{1}^\HH\bw_1|^2}{\sigma^2}
\!\!+\!\!\frac{|\bh_{21}^\HH\bw_{21}|^2}{|\bh_{21}^\HH\bw_{22}|^2\!\!+\!\!\sigma^2}\!\right)\!\!,\!\right.\nonumber\\
&\left.\log_2\!\left(1\!\!+\!\!\frac{\|\bH_0\bw_1\|^2}{\sigma^2}\!\right)\!\vphantom{\log_2\!\left(1\!+\!\frac{|\bh_{1}^\HH\bw_1|^2}{\sigma^2}
\!+\!\frac{|\bh_{21}^\HH\bw_{21}|^2}{|\bh_{21}^\HH\bw_{22}|^2\!+\!\sigma^2}\right)}\right].
\end{align}

For the case that Tu$_2$ helps the transmission of Tu$_1$, if the rate achieved at Ru$_2$ in $t_1$, $R_{12}(\bw_1)$, is greater than $Q_{_{\mathrm{Tu}_1}}(\bw_1,\bw_{21},\bw_{22})$, i.e., $R_{12}(\bw_1)=\frac{1}{2}\log_2\left(1+\frac{|\bh_{12}^\HH\bw_1|^2}{\sigma^2}\right)\geq Q_{_{\mathrm{Tu}_1}}(\bw_1,\bw_{21},\bw_{22})$, Ru$_2$ can decode $x_1$ in $t_1$ and thus, Ru$_2$ can employ SIC to eliminate the effect of $x_1$ in its received signal in $t_2$. Therefore, the achievable rate at Ru$_2$ is given by
\begin{align}
\!R_{_{\mathrm{Ru}_2}}\!\!(\!\bw_{21}\!,\!\!\bw_{22}\!)\!\!=\!\!\!\left\{\!\!\begin{array}{ll}
                       \!\!\!R_{_{\mathrm{Ru}_2}}^{\text{SIC}}(\bw_{22}), \!\text{if}\!\left\{\begin{array}{ll}
                        \!\!\!\!R_{12}(\!\bw_1\!)\!\geq\!Q_{_{\mathrm{Tu}_1}}\!(\bw_1\!,\!\bw_{21}\!,\!\bw_{22}\!)\\ \!\!\!\!\lambda_{\max}(\bH_0^\HH\bH_0)\!>\!\tilde{g}_1
                        \end{array}\right.\!\!\!\!,\!\!\!\!\\
                       \!\!\!R_{_{\mathrm{Ru}_2}}^{\text{NSIC}}\!(\!\bw_{21}\!,\!\bw_{22}\!), \!\text{if}\!\left\{\begin{array}{ll}
                        \!\!\!\!R_{12}\!(\!\bw_1\!)\!\!<\!\!Q_{_{\mathrm{Tu}_1}}\!\!(\!\bw_1\!,\!\bw_{21}\!,\!\bw_{22}\!)\\
                        \!\!\!\!\lambda_{\max}(\bH_0^\HH\bH_0)\!>\!\tilde{g}_1
                        \end{array}\right.\!\!\!\!\!\!,\!\!\!\\
                       \!\!\!\frac{1}{2}\log_2(1+\rho_2\tilde{g}_2),\!\text{otherwise},
                     \end{array}\right.
\end{align}
where $\tilde{g}_2=\|\bh_2\|^2$, $R_{_{\mathrm{Ru}_2}}^{\text{SIC}}(\bw_{22})$ and $R_{_{\mathrm{Ru}_2}}^{\text{NSIC}}(\bw_{21},\bw_{22})$ are given in \eqref{R2-NSIC-SIC-MISO}.

For the MIMO case, in order to maximize the expected achievable rate at Ru$_1$ while guaranteeing QoS requirement at Ru$_2$, the beamformer at Tu$_1$, $\bw_1$, and the beamformers at Tu$_2$, $\bw_{21}$ and $\bw_{22}$,
are jointly optimized by the following problem
\begin{subequations}\label{OP2}
\begin{align}
\mathbf{P4}~:~\max_{\bw_1,\bw_{21},\bw_{22}} &~ R_{_{\mathrm{Ru}_1}}\!(\bw_1,\bw_{21},\bw_{22})\\
\mathrm{s.t.} & ~R_{_{\mathrm{Ru}_2}}(\bw_{21},\bw_{22})\geq Q,&\\
&~ \bw_1^\HH\!\bw_1\!\!\leq\!\! P_1,\bw_{21}^\HH\!\bw_{21}\!\!+\!\!\bw_{22}^\HH\!\bw_{22}\!\!\leq\!\!P_2,\!\!\!
\end{align}
\end{subequations}
where $Q\in[0,Q^{\max}]$. Here, when $Q=Q^{\max}$, Tu$_2$ does not have the residual power to help Tu$_1$ and hence,
we have $\bw_{21}^\HH\bw_{21}=0$ and $\bw_{22}^\HH\bw_{22}=P_{2}$.
The power allocation at Tu$_2$ is embedded in the beamformer design.

Similar with SIMO case, $\mathbf{P4}$ in \eqref{OP2} can be divided into four subproblems with respect to whether Ru$_2$ applies SIC, and the forms of $R_{_{\mathrm{Ru}_1}}\!(\bw_1,\bw_{21},\bw_{22})$. Due to the space limitation and the similarities of the subproblems in MIMO case, in the following, we consider the subproblem $\mathbf{P4-21}$, where
Ru$_2$ does not apply SIC in $t_2$ and
$Q_{_{\mathrm{Tu}_1}}(\bw_1,\bw_{21},\bw_{22})$ is
determined by
\begin{align}\label{QTu1_min}
\!Q_{_{\mathrm{Tu}_1}}\!\!(\!\bw_1\!,\!\bw_{21}\!,\!\bw_{22}\!)\!\!=\!\!\frac{1}{2}\!\log_2\!\left(\!\!1\!\!+\!\!\frac{|\bh_{1}^\HH\!\bw_1|^2}
{\sigma^2}\!\!+\!\!\frac{|\bh_{21}^\HH\!\bw_{21}|^2}{|\bh_{21}^\HH\!\bw_{22}|^2\!\!+\!\!\sigma^2}\right)\!\!.\!\!\!\!\!\!
\end{align}
Therefore, for this case, the subproblem $\mathbf{P3-21}$ is represented by
\begin{subequations}\label{SOP}
\begin{align}
\!\!\!\mathbf{\!P\!4\!\!-\!\!21\!}\!:\!\!\!\!\!\!\!\!\max_{\!\bw_1\!,\!\bw_{21}\!,\!\bw_{22}}\! &\!\!\!\!\!\alpha\!\log_2\!\!\!\left(\!\!\!1\!\!+\!\!\frac{\!\!|\!\bh_{1}^\HH\!\bw_1\!|\!^2\!\!}{\sigma^2}
\!\!+\!\!\frac{|\bh_{21}^\HH\bw_{21}|^2}{\!|\bh_{21}^\HH\!\bw_{22}\!|^2\!\!+\!\!\sigma^2}\!\!\!\right)\!
\!\!+\!\!(\!1\!\!-\!\!\alpha\!)\!\!\log_2\!\!\!\left(\!\!\!1\!\!+\!\!\frac{\!|\!\bh_{1}^\HH\!\bw_1\!|\!^2\!\!}{\sigma^2}\right)\!&\!\!\label{SOP-OF}\\
\!\!\mathrm{s.t.}
\!\log_2\!&\!\left(\!\!1\!\!+\!\!\frac{|\bh_{12}^\HH\!\bw_1\!|^2}{\sigma^2}\!\!\right)\!\!\!<\!\! \log_2\!\!\left(\!\!\!1\!\!+\!\!\frac{|\bh_{1}^\HH\!\bw_1\!|^2}{\sigma^2}\!\!+\!
\!\frac{\!|\bh_{21}^\HH\!\bw_{21}\!|\!^2}{\!|\bh_{21}^\HH\!\bw_{22}\!|^2\!\!+\!\!\sigma^2}\!\!\!\right)\!\!,\!\!\!&\label{SOP-C1}\\
\!\log_2\!&\!\left(\!\!\!1\!\!+\!\!\frac{\|\!\bH_0\!\bw_1\!\|^2}{\sigma^2}\!\right)\!\!\!\geq \!\! \log_2\!\!\!\left(\!\!\!1\!\!+\!\!\frac{|\bh_{1}^\HH\!\bw_1\!|^2}{\sigma^2}\!\!+\!
\!\frac{|\bh_{21}^\HH\!\bw_{21}\!|^2}{|\bh_{21}^\HH\!\bw_{22}\!|^2\!\!+\!\!\sigma^2}\!\!\!\right)\!\!,\!\!&\label{SOP-C2}\\
\!\frac{1}{2}\!\!\log_2\!\!&\left(1\!+\!\frac{|\bh_2^\HH\bw_{22}|^2}{|\bh_2^\HH\bw_{21}|^2+
\sigma^2}\right)\geq Q,&\label{SOP-C3}\\
\!\bw_1^\HH~&\!\!\!\bw_1\leq P_1,~\bw_{21}^\HH\bw_{21}+\bw_{22}^\HH\bw_{22}\leq P_2,&\label{SOP-C4}
\end{align}
\end{subequations}
where the constraint \eqref{SOP-C1} is the condition that Ru$_2$ cannot apply SIC in $t_2$
and the constraint \eqref{SOP-C2} is the condition that \eqref{Q_t_MIMO} becomes \eqref{QTu1_min}.
Since $\mathbf{P4-21}$ is non-convex and $\bw_1$, $\bw_{21}$ and $\bw_{22}$ are coupled in the constraints, it is hard to directly obtain the solution of $\mathbf{P4-21}$.

In order to obtain the beamformers for MIMO case, we combine the optimal beamformer structure of MISO case, which is given in Lemma \ref{lemma_ryu}, to obtain $\bw_1$,
and the BCU based iterative algorithm for SIMO case, which is given in TABLE \Rmnum{1}, to obtain $\bw_{21}$ and $\bw_{22}$.
At one iteration, for given $\bw_{21}$ and $\bw_{22}$, we obtain $\bw_1$ based on the optimal structure in \eqref{w_op}. Then, at the next iteration,
$\bw_{21}$ and $\bw_{22}$ are optimized for the fixed $\bw_{1}$ by using BCU based algorithm to maximize the expected achievable rate.

For given $(\bw_{21}, \bw_{22})$, the terms related to $(\bw_{21}, \bw_{22})$ can be regarded as the constant and then, we can see that the problem $\mathbf{P4-21}$ has the same structure of the problem
for MISO case, which optimizes $\bw_{1}$ only.
Hence, for given $(\bw_{21}, \bw_{22})$, we design $\bw_{1}$ based on the optimal structure of the beamformer for MISO, given in \eqref{w_op}.
The optimal beamformer for MISO is constructed by two bases $\bw_0=\frac{\bPi_{\mathbf{h}_0}\bh_1}{\left\|\bPi_{\mathbf{h}_0}\bh_1\right\|}$ and $\bw_0^\bot=\frac{\bPi_{\mathbf{h}_0}^\bot\bh_1}{\left\|\bPi_{\mathbf{h}_0}^\bot\bh_1\right\|}$.
Here, since $\bh_0=\theta\bPi_{\mathbf{h}_0}\bh_1$ for some scalar $\theta$ and $\bh_1=\bPi_{\mathbf{h}_0}\bh_1+\bPi_{\mathbf{h}_0}^\bot\bh_1$, the optimal beamformer for MISO in \eqref{w_op} can be represented by two bases $\bw_0^{\text{mrt}}=\frac{\bh_0}{\|\bh_0\|}$ and $\bw_1^{\text{mrt}}=\frac{\bh_1}{\|\bh_1\|}$ such as
\begin{align}
\bw_1^{\mathrm{opt}}=\lambda_1\bw_0^{\text{mrt}}+\lambda_2\bw_1^{\text{mrt}},\label{structure2}
\end{align}
where $\lambda_1$ and $\lambda_2$ are determined to satisfy $\|\bw_1^\text{opt}\|^2=1$.
From \eqref{structure2}, we note that the optimal structure of the beamformer for MISO is the linear combination of MRT beamformers of channels $\bh_0$ and $\bh_1$.
Thus, for MIMO case, we design $\bw_1$ based on the beamformer structure in \eqref{structure2} as
\begin{align}
\bw_1(\lambda)=\sqrt{P_1}\frac{\lambda \bw_0^{\text{eig}}+(1-\lambda)\bw_1^{\text{mrt}}}{\|\lambda \bw_0^{\text{eig}}+(1-\lambda)\bw_1^{\text{mrt}}\|}, \label{structure3}
\end{align}
where $\bw_0^{\text{eig}}$ is an eigenvector corresponding the largest eigenvalue of $\bH_0^\HH\bH_0$ and $\lambda$ is a real value in $0\leq \lambda \leq 1$.
For the MIMO channel, the eigenvector corresponding the largest eigenvalue of $\bH_0^\HH\bH_0$ is
the beamformer that maximizes the achievable rate of $\bH_0$, similar to MRT beamformer for MISO channel.
Based on \eqref{structure3}, for given $(\bw_{21}, \bw_{22})$, we optimize the coefficient $\lambda$
to maximize the expected achievable rate at Ru$_1$ by one-dimensional line search.

Similarly, for given $\bw_1$, the terms related to $\bw_1$ can be regarded as the constant and thus, the problem $\mathbf{P4-21}$ becomes the problem that has the same structure of that
for SIMO case, which optimizes $\bw_{21}$ and $\bw_{22}$ only.
Therefore, for given $\bw_1$, we can obtain $\bw_{21}$ and $\bw_{22}$
by using the BCU based iterative algorithm for SIMO, proposed in Section \Rmnum{3}-C.
Consequently, for MIMO case, we obtain the beamforers $\bw_1$, $\bw_{21}$ and $\bw_{22}$
iteratively to maximize the expected achievable rate at Ru$_1$ while guaranteeing the QoS of Ru$_2$.
The details of the proposed algorithm for MIMO case are summarized in TABLE \Rmnum{2}.

\begin{table}
\caption{The Proposed Algorithm For MIMO}
\begin{center}
\begin{tabular}{|l|}
\hline
\textbf{Initialization:} Define $\lambda_m=\frac{m}{M}, m=1,\cdots,M$, where $M$\\
\qquad\qquad\qquad\ is a positive integer. Generate feasible block\\
\qquad\qquad\qquad\ variables $(\bw_{21}^0,\bw_{22}^0)$.\\
\textbf{Repeat:} \textbf{\textit{For}} each given $\lambda_m$, set\\
\qquad\quad $\bw_1\!\!:=\!\!\bw_1(\!\lambda_m\!)\!\!=\!\!\sqrt{P_1}\frac{\!\lambda_m\! \bw_0^{\text{eig}}\!+\!(1\!-\!\lambda_m)\bw_1^{\text{mrt}}}
{\|\lambda_m\bw_0^{\text{eig}}+(1-\lambda_m)\bw_1^{\text{mrt}}\|}$, and do\\
\qquad\quad S1--S3 to search for $(\!\bw_{21,\lambda_m}^*\!,\! \bw_{22,\lambda_m}^*\!)$ related with $\lambda_m$.\\
~~~~\text{S1}: \textbf{\textit{For}} $k=1,\cdots,K$, where $K$ is the maximal iteration\\
~~~~\quad\qquad\ \ times, do the following until converge.\\
~~~~\text{S2}: Solve problem $\mathbf{\!P4\!-\!21\!}$ without constraints \eqref{SOP-C1}\eqref{SOP-C2}\\
~~~\quad\quad based on S2--S5 in TABLE \Rmnum{1}.\\
  \hline
\end{tabular}
\end{center}
\end{table}

\section{Simulation Results}\label{simulation results}
In this section, we evaluate the performance of the trust degree based user cooperation for three cases:
SISO case where all users have a single antenna ($N_1=N_2=1$), MISO case where Tu$_1$ has the multiple antennas ($N_1=2,\ N_2=1$), SIMO case where Tu$_2$ has the multiple antennas ($N_1=1,\ N_2=2$) and MIMO case where both Tu$_1$ and Tu$_2$ have the multiple antennas ($N_1=N_2=2$).
Unless otherwise specified, we use the average gains of channel elements as
$\left\{\sigma^2_{H_0},\sigma^2_{h_1},\sigma^2_{h_2},\sigma^2_{h_{12}},
\sigma^2_{h_{21}}\right\}=\left\{-35,-45,-30,-25,-25\right\}$dB and the expected achievable rates
are averaged over $10^4$ channel realizations.

\subsection{SISO case ($N_1=N_2=1$)}
\begin{figure}
\centering
 \includegraphics[width=1\columnwidth]{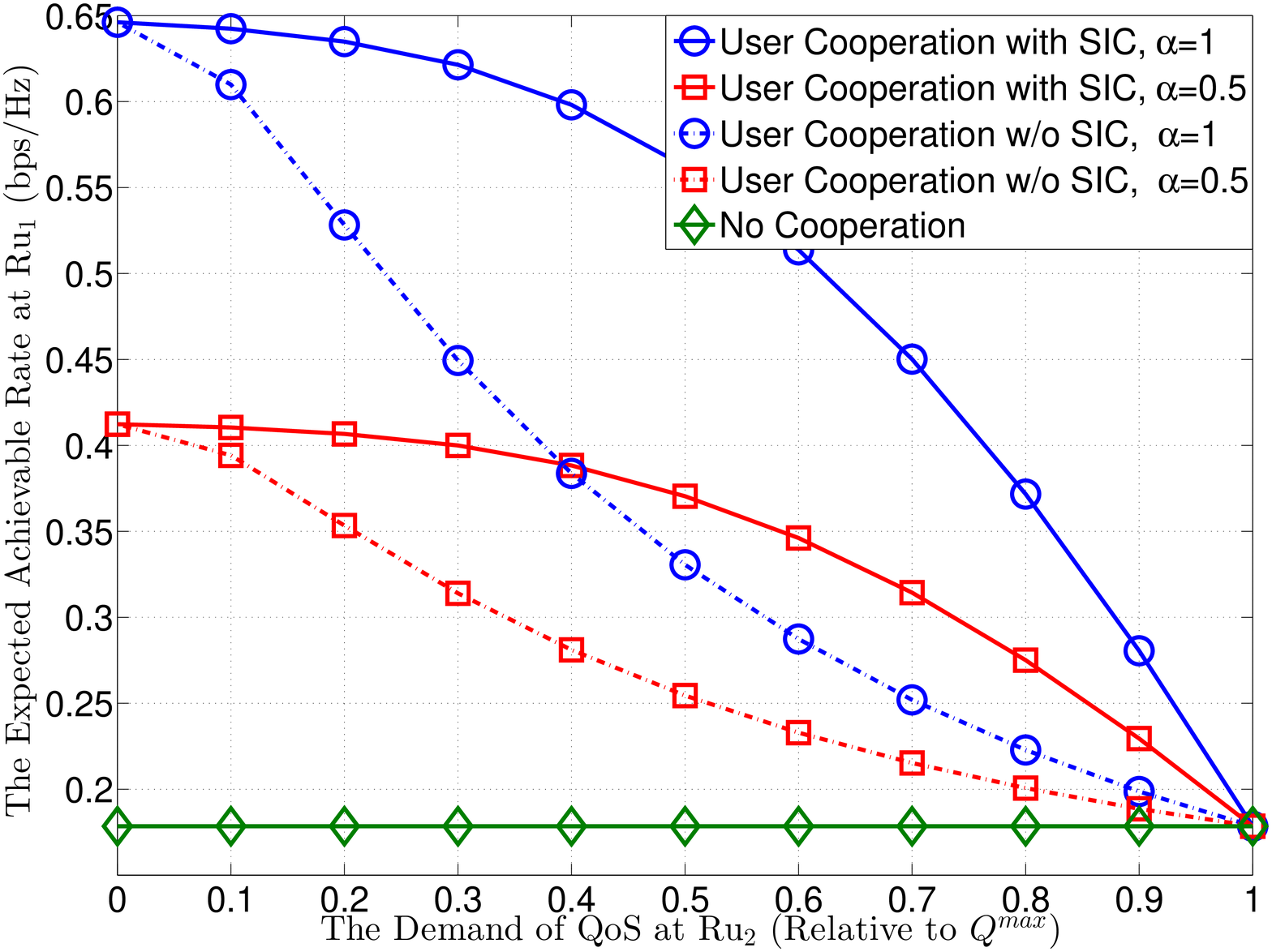}\\
   \caption{The expected achievable rate at $\text{Ru}_1$ versus the QoS requirement at Ru$_2$, where $\rho_1=\rho_2=40\text{dB}$ and $N_1=N_2=1$.}\label{2}
\end{figure}

\begin{figure}
\centering
  \includegraphics[width=1\columnwidth]{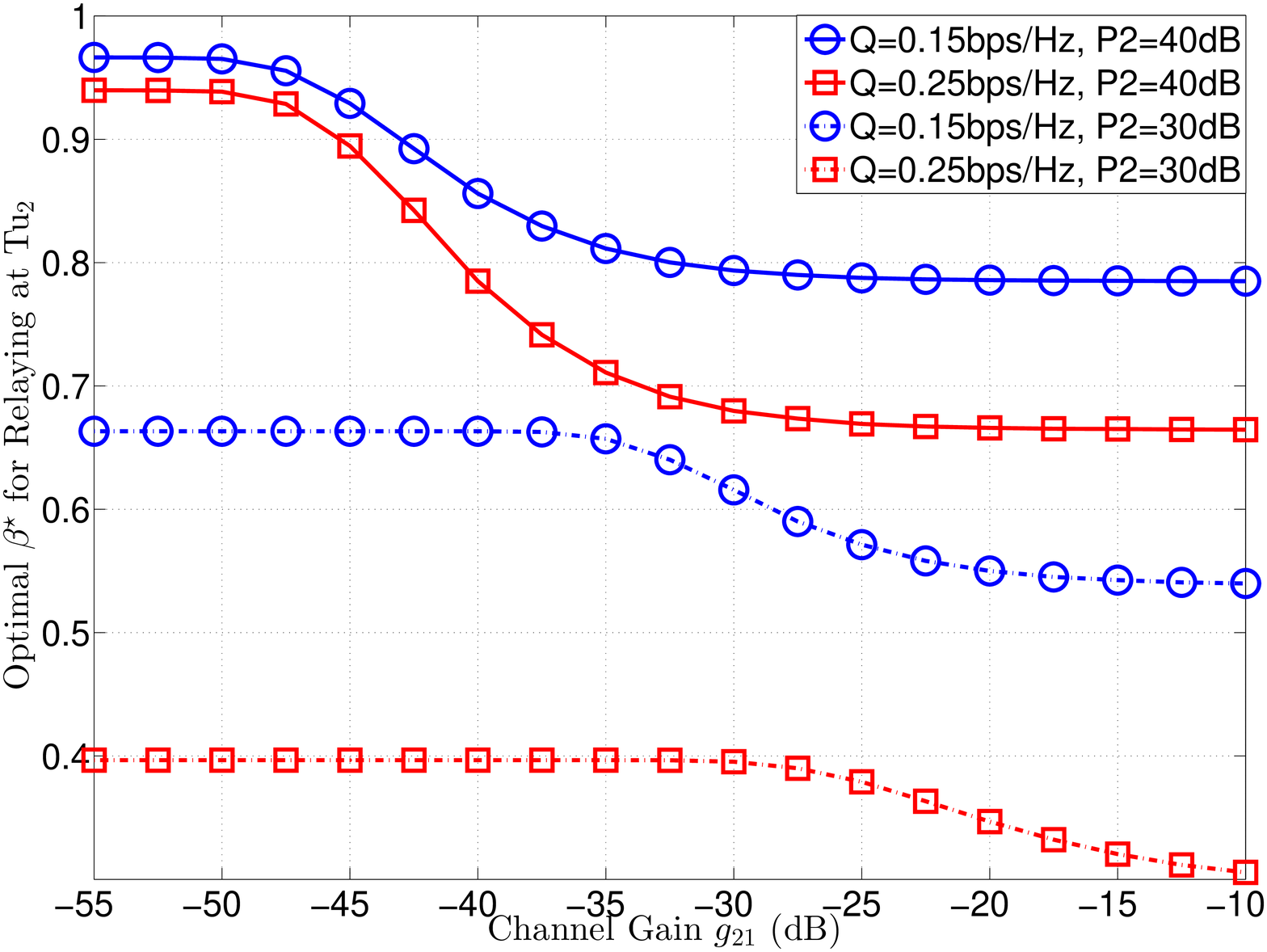}\\
   \caption{Optimal $\beta^\star$ at $\text{Tu}_2$ versus channel gain $g_{21}$ with different given QoS and $P_2$, $\alpha=0.5$, $\rho_1=40\text{dB}$ and $N_1=N_2=1$.}\label{3}
\end{figure}

In Fig.\ref{2}, for SISO case, we plot the expected achievable rates of Ru$_1$ according to the QoS requirement at Ru$_2$, $Q$, when the transmit SNR at Tu$_1$ and Tu$_2$ are given by $\rho_1=\rho_2=40$dB.
To compare with the proposed user cooperation scheme, which applies SIC, we also plot the expected achievable rates of the user cooperation without SIC and no cooperation ($\alpha=0$).
For the proposed user cooperation, the optimal power allocation at Tu$_2$ for cooperation is obtained as $\beta^{\star}$ in Theorem \ref{theorem}.
In Fig.\ref{2}, we show that the expected achievable rate can be significantly increased by the user cooperation when the trust degree between users is high such as $\alpha=1$.
For the case with $\alpha=1$, Tu$_2$ always helps the transmission of Tu$_1$ while
when $\alpha=0.5$, Tu$_2$ helps the transmission of Tu$_1$ with probability $0.5$
even if Tu$_2$ has sufficient power budget after achieving QoS.
In addition, when the QoS requirement at Ru$_2$ is small,
Tu$_1$ can achieve very high expected achievable rate because Tu$_2$ has a large amount of residual power after achieving its QoS and helps the transmission of Tu$_1$ by using large power.
By applying SIC at Ru$_2$, the expected achievable rate at Ru$_1$ can be further improved
since Tu$_2$ can achieve QoS requirement with small power, and hence Tu$_2$ can allocate more power for cooperation than that without SIC.

In Fig.\ref{3}, for different $Q$ ($Q=0.5$ and $0.3$bps/Hz) and $\rho_2$ ($\rho_2=40$ and $30$dB), the optimal power allocation for cooperation at Tu$_2$, $\beta^{\star}$, is plotted as a function of the channel gain from Tu$_2$ to Ru$_1$, $g_{21}$,
when $\alpha=0.5$ and $\rho_{1}=40$dB.
In this figure, we use the channel gains as $\{\sigma^2_{H_0},\sigma^2_{h_1},\sigma^2_{h_2},\sigma^2_{h_{12}}\}=\{-32,-40,-30,-32\}$dB.
When $g_{21}$ is weak, from \eqref{Q_Tu}, Ru$_2$ can apply SIC and Tu$_2$ can allocate more power for cooperation. Hence, the optimal power allocation is obtained by $\beta^{\star}=\beta_{Q_1}$.
In contrast, when $g_{21}$ is strong, Tu$_2$ cannot allocate large power for cooperation
and the optimal power allocation is obtained by $\beta^{\star}=\beta_{Q_2}$.
Otherwise, Tu$_2$ reduces $\beta$ to help applying SIC at Ru$_1$, and
the optimal power allocation is determined by ${\beta}^{\star}=\tilde{\beta}_{1}$, which is a decreasing function of $g_{21}$.
In addition, we can see that Tu$_2$ can allocate more power for cooperation
when the QoS requirement at Ru$_2$, $Q$, is small or the amount of transmit power budget at Tu$_2$ is large.

\subsection{MISO case ($N_1=2, \ N_2=1$)}
In the MISO case, we jointly design the beamformer at Tu$_1$, $\bf{w}_1$, and the power allocation for cooperation at Tu$_2$, $\beta$, according to trust degree, $\alpha$.
For the proposed user cooperation, the beamformer at Tu$_1$ is obtained by Theorem \ref{theorem2}, and
the corresponding power allocation is obtained by one dimensional search.
For the comparison, we show the performance of the case that Tu$_1$ simply uses a MRT beamformer, $\!\bf{w}_1\!=\!\bf{w}_1^{\mathrm{mrt}}\!=\!\frac{\bf{h}_1}{\|\bf{h}_1\|}\!$, and Ru$_2$ does not apply SIC. The performance of the no cooperation between users $(\!\alpha\!=\!0\!)$ with MRT beamformer is also given as a baseline.
\begin{figure}
\centering
  \includegraphics[width=1\columnwidth]{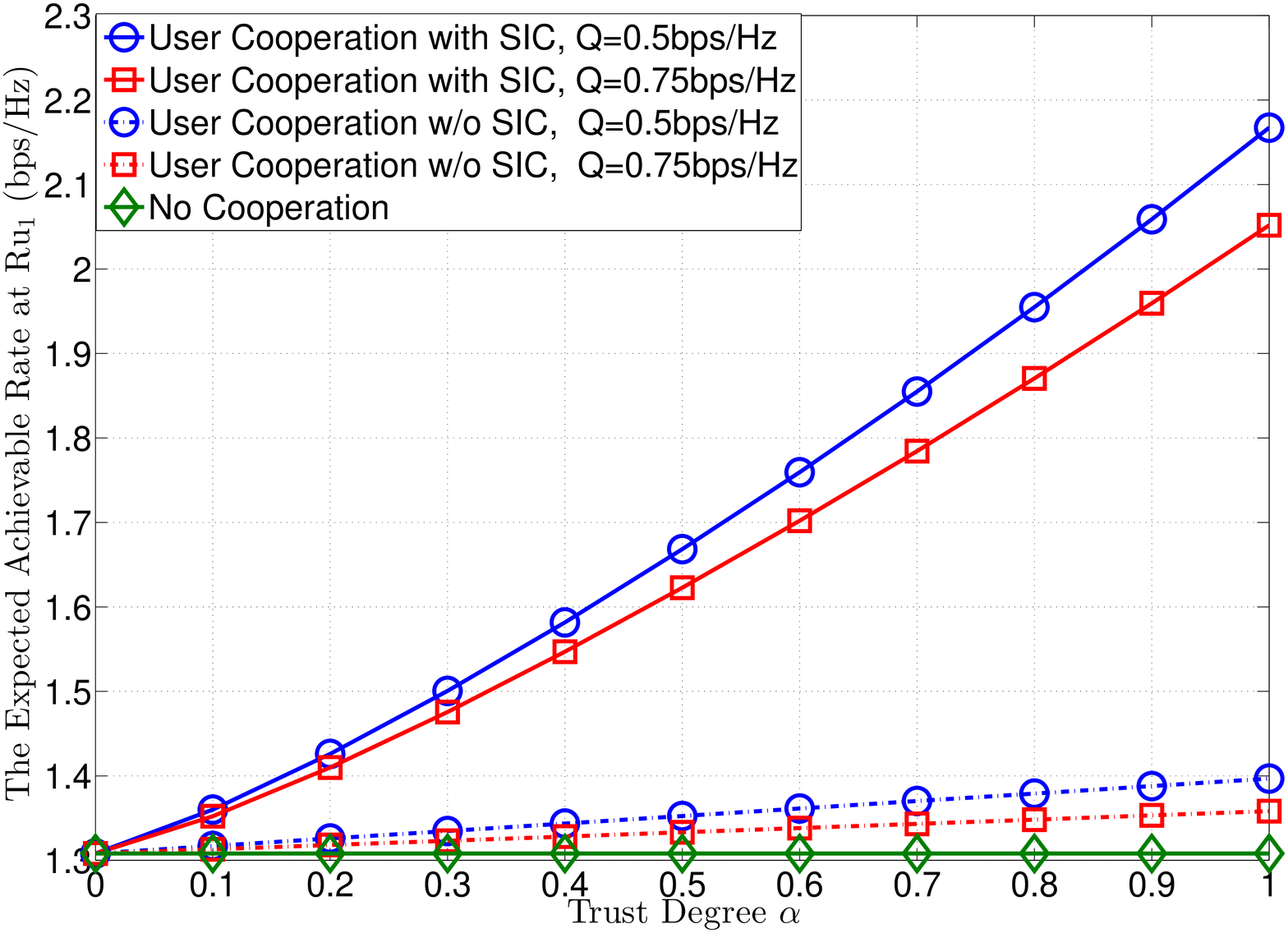}\\
   \caption{The expected achievable rate at $\text{Ru}_1$ versus trust degree $\alpha$, where $\rho_1=\rho_2=50\text{dB}$ and $N_1=2,\ N_2=1$.}\label{4}
\end{figure}

\begin{figure}
\centering
  \includegraphics[width=1\columnwidth]{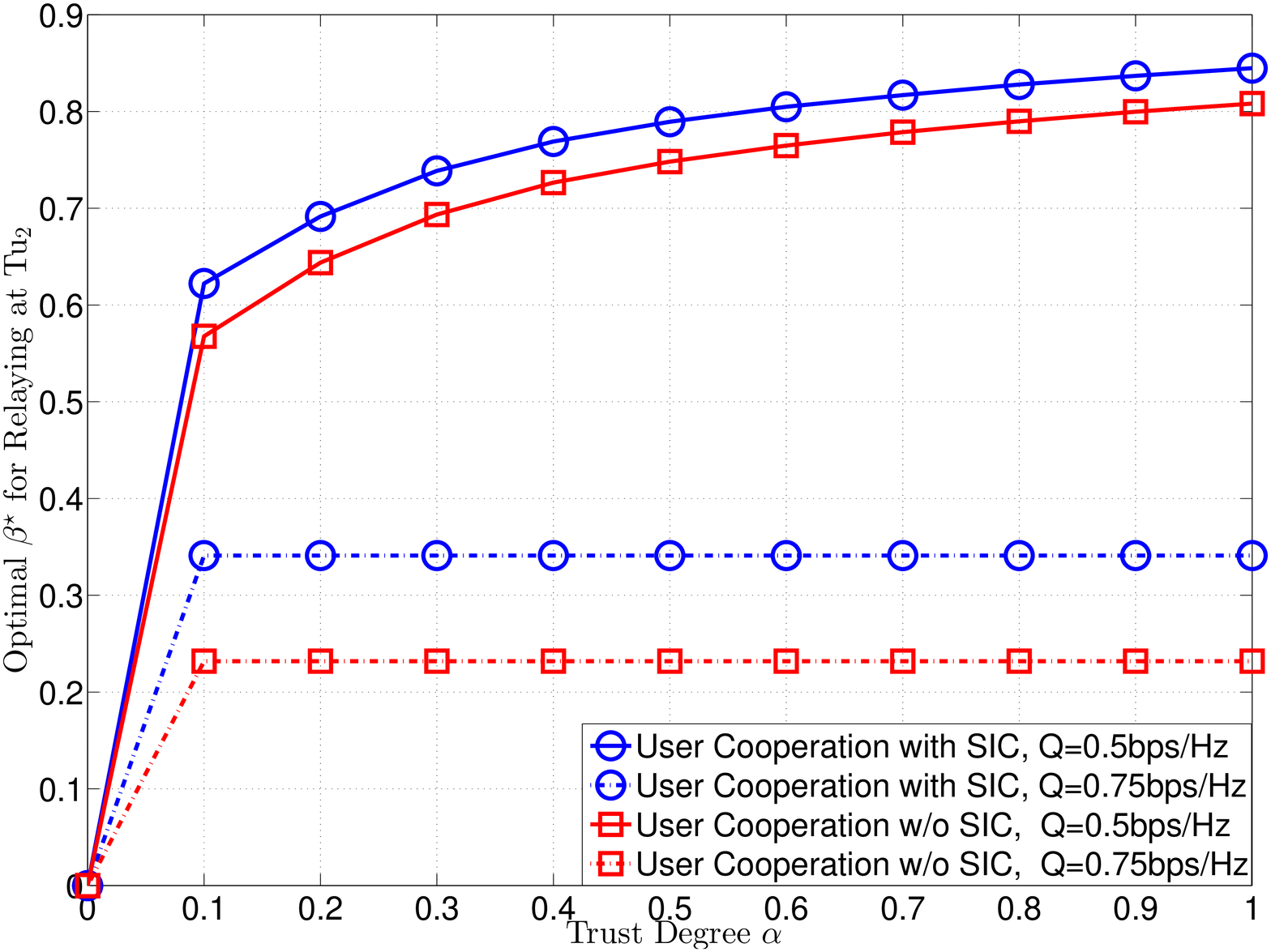}\\
   \caption{Optimal $\beta^\star$ for relaying at $\text{Tu}_2$ versus trust degree $\alpha$, where $\rho_1=\rho_2=50\text{dB}$ and $N_1=2,\ N_2=1$.}\label{5}
\end{figure}

In Fig. \ref{4}, for the proposed and reference schemes, we plot the expected achievable rates of Ru$_1$ according to trust degree, $\alpha$ when $\!\rho_1\!=\!\rho_2\!=\!50$dB.
With the growth of $\alpha$, for the user cooperation schemes,
the expected achievable rates of Ru$_1$ are increased by the cooperation of Tu$_2$.
For the proposed user cooperation, since the beamformer is efficiently designed by considering both trust degree and physical channel qualities, the performance improvement becomes significantly large according to $\alpha$.
However, when Tu$_1$ uses MRT beamformer, the expected achievable rate improvement is marginal because beamformer is designed independently from $\alpha$ and hence, the benefit from the cooperation cannot be fully exploited even the trust degree is high.

Fig. \ref{5}, the optimal power allocation for cooperation at Tu$_2$, $\beta^{\star}$, is plotted according to $\alpha$. When the Tu$_1$ transmits its data with MRT beamformer, which is independently designed from $\alpha$,
the corresponding power allocation to maximize the expected achievable rate is also determined independently as a  constant.
For the proposed beamforming, in Fig. \ref{5}, we can see that $\beta^{\star}$ increases with the growth of trust degree $\alpha$. In the proposed beamforming, when $\alpha$ is high, the direction of beamformer
is steered from $\mathbf{h}_{1}$ to $\mathbf{h}_{0}$ to fully utilize the cooperation of Tu$_2$.
Due to constraint of DF relaying, the expected achievable rate in \eqref{Q_t_MISO_1} is maximized by balancing the minimum rates achieved at Ru$_1$ (first term) and Tu$_2$ (second term).
Hence, for high $\alpha$, \eqref{Q_t_MISO_1} is maximized by increasing the second term from beamforming design
and compensating the first term by assigning large power for cooperation at Tu$_2$, i.e., $\beta^{\star}$ is high.

\subsection{SIMO case ($N_1=1, N_2=2$) and MIMO case ($N_1=2, N_2=2$)}
For the SIMO case, we evaluate the performance of the proposed user cooperation based on trust degree in terms of the expected achievable rate. The expected achievable rate at the proposed scheme is achieved by beamformers obtained from the proposed algorithm, which is given in TABLE \Rmnum{1}. In Fig. \ref{6}, for $Q=0.5$ and $1$bps/Hz, we plot the expected achievable rates of the proposed and reference schemes versus the trust degree, $\alpha$ when $\rho_1=\rho_2=50$dB. Similar to previous subsections, we can see that for high $\alpha$,
the expected achievable rate at Ru$_1$ is significantly increased by the cooperation with Tu$_2$
and the performance is further improved by efficiently designing beamformers based on the proposed algorithm. For the MIMO case, we can see the similar phenomena for $Q=1$ and $2$bps/Hz in Fig. \ref{8}.

In Fig. \ref{7}, for $Q=0.5$ and $1$bps/Hz, the expected achievable rate at Ru$_1$ is plotted according to the relaying channel quality from Tu$_2$ to Ru$_1$, $\tilde{g}_{21}=\|\bh_{21}\|^2$. In this figure, the trust degree and transmit SNRs are $\alpha=0.5$ and  $\rho_1=\rho_2=50$dB.
From Fig. \ref{7}, we first observe that when the gain of relaying channel increases from $-50$dB to $-25$dB, the expected achievable rate at Ru$_1$ can be increased by cooperative transmission from Tu$_2$ via $\bh_{21}$.
However, the expected achievable rate does not increase and is saturated when $\tilde{g}_{21}$ increases in the regime of $\tilde{g}_{21}>-25$dB. Since when the quality of the relaying channel, $\bh_{21}$, is much better than that of channel between Tu$_1$ and Tu$_2$, $\mathbf{h}_{0}$, due to the DF relaying constraint,
the rate achieved at Tu$_2$, which is the second term of \eqref{Q_t_SIMO}, is always lower than
the rate achieved at Ru$_1$, which is the first term of \eqref{Q_t_SIMO} for all feasible beamformers.
Hence, the expected achievable rate cannot increase and is saturated even if the quality of the relaying channel is sufficiently good. For the MIMO case, we can see the similar phenomena for $Q=1$ and $2$bps/Hz in Fig. \ref{9}.
\begin{figure}
\centering
  \includegraphics[width=1\columnwidth]{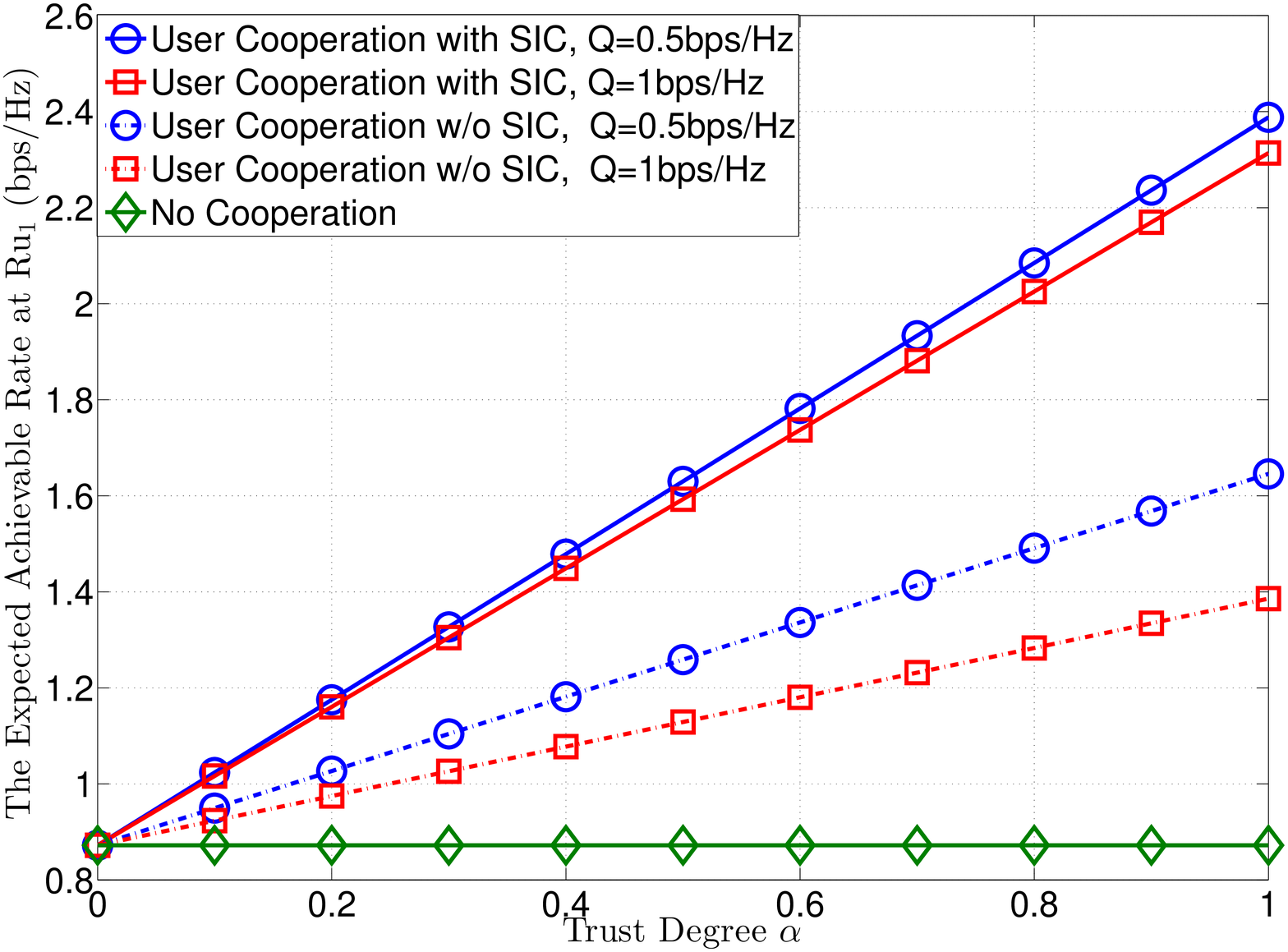}\\
   \caption{The expected achievable rate at $\text{Ru}_1$ versus trust degree $\alpha$, where $\rho_1=\rho_2=50\text{dB}$ and $N_1=1, N_2=2$.}\label{6}
\end{figure}

\begin{figure}
\centering
 \includegraphics[width=1\columnwidth]{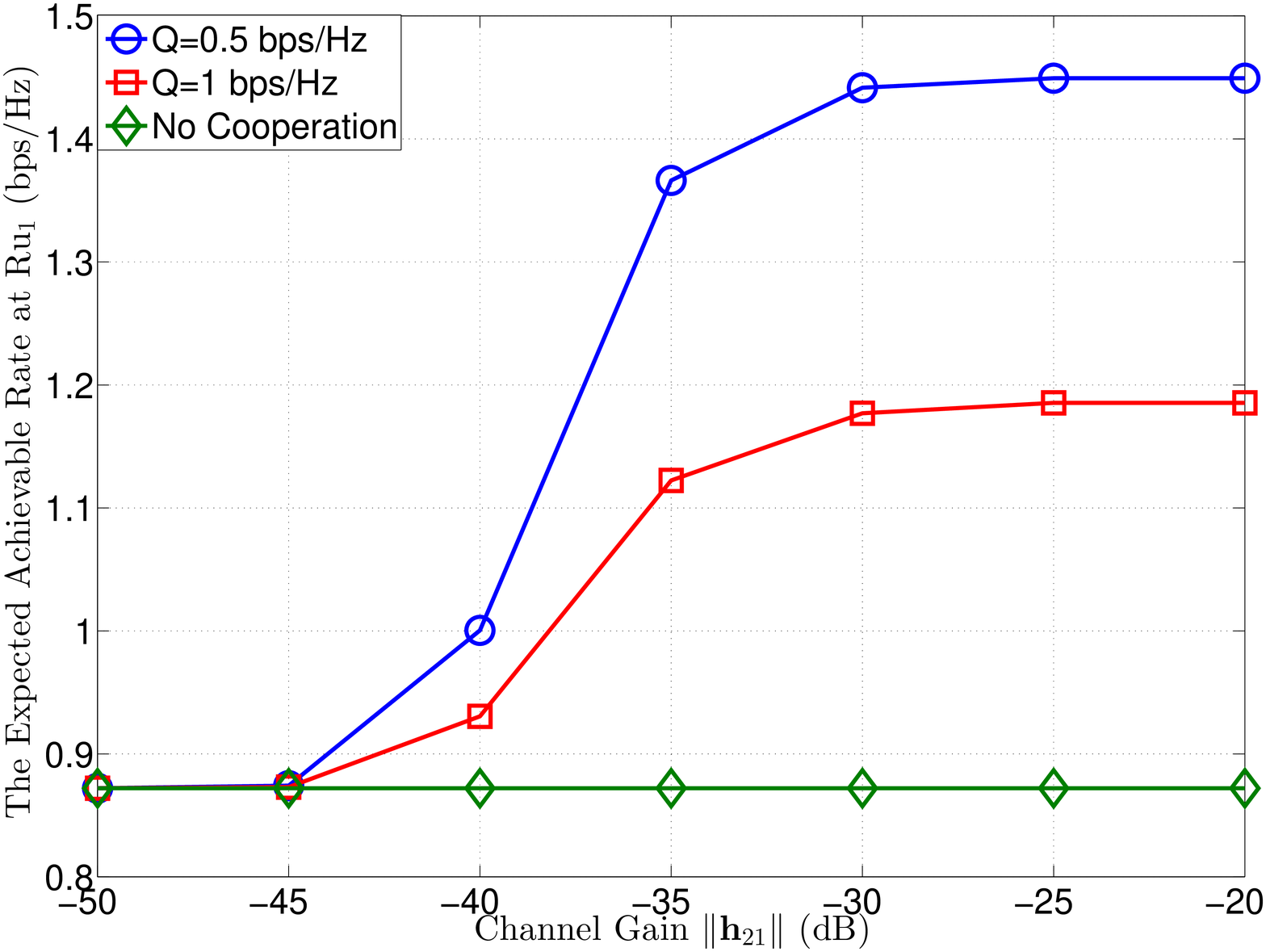}\\
   \caption{The expected achievable rate at $\text{Ru}_1$ versus channel gain $\|\bh_{21}\|^2$ with respect to different given QoS, where $\rho_1=\rho_2=50\text{dB}$, $N_1=1, N_2=2$ and $\alpha=0.5$.}\label{7}
\end{figure}

\begin{figure}
\centering
  \includegraphics[width=1\columnwidth]{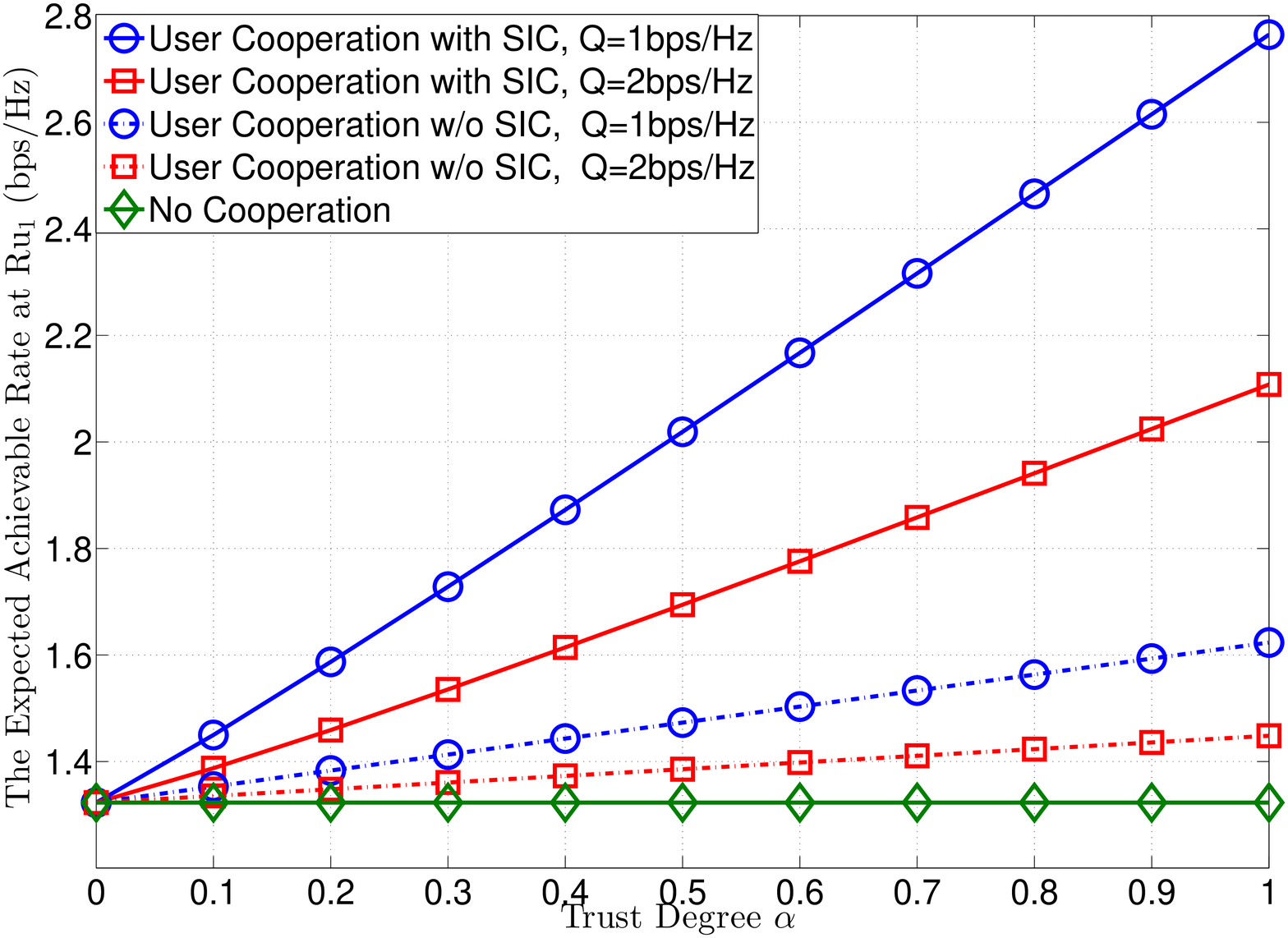}\\
   \caption{The expected achievable rate at $\text{Ru}_1$ versus trust degree $\alpha$, where $\rho_1=\rho_2=50\text{dB}$ and $N_1=N_2=2$.}\label{8}
\end{figure}

\begin{figure}
\centering
  \includegraphics[width=1\columnwidth]{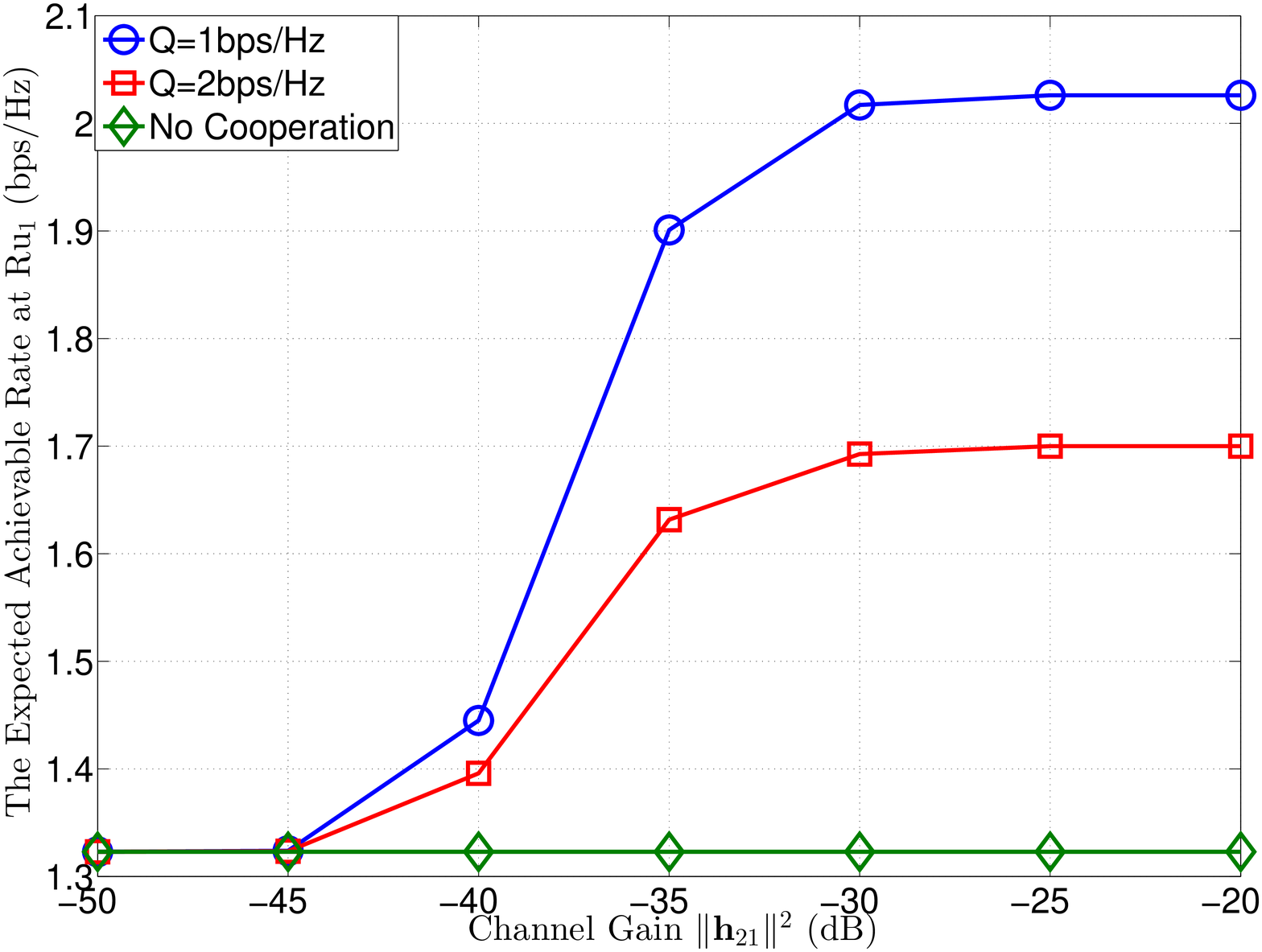}\\
   \caption{The expected achievable rate at $\text{Ru}_1$ versus channel gain $\|\bh_{21}\|^2$ with different given QoS, where $\rho_1=\rho_2=50\text{dB}$, $N_1=N_2=2$ and $\alpha=0.5$.}\label{9}
\end{figure}

\section{Conclusion}\label{conclusion}
In this paper, we propose the user cooperation techniques in the multiple antenna system with
two communication pairs, i.e., Tu$_1$-Ru$_1$ and Tu$_2$-Ru$_2$, where Tu$_2$ can help the transmission of Tu$_1$ according to the trust degree.
For different antenna configurations at Tu$_1$ and Tu$_2$, we design the user cooperation strategies by taking into account the trust degree information as well as channel information.
For the SISO case, as a special case, we first propose an optimal power allocation strategy at Tu$_2$, which maximizes
the expected achievable rate at Ru$_1$ while guaranteeing QoS requirement at Ru$_2$, according to the channel qualities and QoS requirement.
For the MISO case, we provide an optimal structure of beamformer as a linear combination of the weighted channel vectors. Then, based on the optimal structure, we obtain the beamformer that maximizes an approximated expected achievable rate as a function of the trust degree and corresponding power allocation at Tu$_2$. For the SIMO case, to jointly optimize the beamformers of Tu$_2$, we utilize semidefinite relaxation (SDR) technique and block coordinate update (BCU) method to solve the considered problem, and guarantee the rank-one solutions at each step. Furthermore, for the MIMO case, the similarities among problem structures related to MISO, SIMO and MIMO cases inspire us to combine the design of beamformer at Tu$_1$ from MISO and the alternative algorithm from SIMO together to jointly optimize the beamformers at Tu$_1$ and Tu$_2$ to maximize the expected achievable rate at Ru$_1$.
Finally, we show that the trust degree between users can be used to
significantly improve the expected achievable rate in the user cooperation networks.

\appendices
\def\thesection{\Alph{section}}%
\def\thesectiondis{\Alph{section}}%

\section{Proof of Theorem \ref{theorem}}\label{app1}
\setcounter{equation}{0}
\renewcommand{\theequation}{A.\arabic{equation}}
From \eqref{R1}, we note that the expected achievable rate at Ru$_1$, $R_{_{\mathrm{Ru}_1}}\!(\beta)$, is
an increasing function with $\beta$.
Hence, the optimal power allocation is determined by the maximum $\beta$ that satisfies the QoS requirement at Ru$_2$ such as $R_{_{\mathrm{Ru}_2}}\!(\beta)\geq Q$.
If SIC can be applied at Ru$_2$, $R_{_{\mathrm{Ru}_2}}\!(\beta)$ is given by $R_{_{\mathrm{Ru}_2}}^\mathrm{SIC}(\beta)$
and otherwise, $R_{_{Ru_2}}\!(\beta)$ is given by $R_{_{\mathrm{Ru}_2}}^\mathrm{NSIC}(\beta)$.
In order to find maximum $\beta$ that satisfies $R_{_{\mathrm{Ru}_2}}\!(\beta)\geq Q$, we first find the conditions that SIC can be applied at Ru$_2$ and for these conditions, we find the optimal $\beta$.

From \eqref{min}, $Q_{_{\mathrm{Tu}_1}}\!(\beta)$ can be represented by two cases of $\beta_{0}\leq\beta\leq 1$ and ${0}\leq\beta\leq \beta_{0}$.
First, for the case of $\beta_{0}\leq\beta\leq 1$, $Q_{_{\mathrm{Tu}_1}}\!(\beta)$ is given by
\begin{align}
Q_{_{\mathrm{Tu}_1}}=\frac{1}{2}\log_2\left(1+\rho_1g_0\right),
\end{align}
which is a constant and independent from $\beta$.
For this case, if $g_{12}\!\!\geq\!\!g_0$, we have $R_{12}\!\!\geq\!\!Q_{_{\mathrm{Tu}_1}}$ and thus, Ru$_2$ can apply SIC. To satisfy QoS requirement at Ru$_2$, we obtain the condition of $\beta$ as
\begin{align}
R_{_{\mathrm{Ru}_2}}^\mathrm{SIC}(\beta)\geq Q~\Rightarrow~\beta\leq\beta_{Q_1},\label{betaQ1cond0}
\end{align}
where $\beta_{Q_1}$ is given in \eqref{betaQ1}.
In this case, the optimal $\beta$ is feasible as $\beta^\star=\beta_{Q_1}$ if $\beta_{0}\leq\beta_{Q_1}$,
and thus, the condition of $Q$ that makes $\beta^\star$ feasible can be obtained by
\begin{align}
\beta_{0}\leq\beta_{Q_1}~\Rightarrow~Q\leq r_{1},\label{betaQ1cond1}
\end{align}
where $r_{1}$ is given in \eqref{rs}.
Therefore, if $g_{12}\geq g_0$ and $Q\leq r_{1}$, the optimal $\beta$ is obtained by $\beta^\star=\beta_{Q_1}$ and QoS of Ru$_2$ is achieved by applying SIC at Ru$_2$.

For the case of ${0}\leq\beta\leq \beta_{0}$, $Q_{_{\mathrm{Tu}_1}}\!(\beta)$ is given by
\begin{align}
Q_{_{\mathrm{Tu}_1}}\!(\beta)=\frac{1}{2}\log_2\left(1\!+\!\rho_1g_1\!+\!\frac{\beta\rho_2g_{21}}{(1\!-\!\beta)\rho_2 g_{21}\!+\!1}\right). \label{QTu11}
\end{align}
For this case, by using \eqref{QTu11}, the condition that can apply SIC at Ru$_2$ is obtained by
\begin{align}
R_{12}\!=\!\frac{1}{2}\log(1+\rho_{1}g_{12})\!\geq\! Q_{_{\mathrm{Tu}_1}}\!(\beta)\!\Rightarrow\!g_{12}\!\geq\! g_{1},\beta\leq\tilde{\beta}_{1},\!\!\!\!\!
\end{align}
where $\tilde{\beta}_{1}$ is given in \eqref{betatilde}.
Thus, if $\beta_{Q_{1}}\leq\beta_{0}$ and $\beta_{Q_{1}}\leq\tilde{\beta}_{1}$,
the optimal $\beta$ is obtained by $\beta^\star=\beta_{Q_{1}}$ because $\beta_{Q_{1}}$ is the maximum $\beta$ that satisfies
the QoS requirement at Ru$_2$ as given in \eqref{betaQ1cond0}.
We obtain the condition of $Q$ that satisfies $\beta_{Q_{1}}\leq\beta_{0}$ and
$\beta_{Q_{1}}\leq\tilde{\beta}_{1}$ as
\begin{align}
\left\{\begin{array}{ll}\beta_{Q_{1}}\leq\beta_{0}&\Rightarrow~Q\geq r_{1},\\
\beta_{Q_{1}}\leq\tilde{\beta}_{1}&\Rightarrow~Q\geq r_{2},
\end{array}\right.
\Rightarrow~Q\geq \max(r_{1},r_{2}),
\end{align}
where $r_{2}$ is given in \eqref{rs}. Therefore, if $g_{12}\geq g_{1}$ and $Q\geq \max(r_{1},r_{2})$,
the optimal $\beta$ is obtained by $\beta^\star=\beta_{Q_{1}}$ and QoS requirement at Ru$_2$ is achieved by applying SIC at Ru$_2$.

On the other hand, if $\tilde{\beta}_{1}\leq\beta_{Q_{1}}$, equivalent $Q\leq r_{2}$,
we cannot guarantee the QoS requirement at Ru$_2$ by $\beta=\beta_{Q_{1}}$ because
Ru$_2$ cannot apply SIC when $\beta=\beta_{Q_{1}}$.
Hence, for $Q\leq r_{2}$, the QoS requirement at Ru$_2$ can be guaranteed by $\beta=\tilde{\beta}_{1}$ with applying SIC
or $\beta=\beta_{Q_{2}}$ without applying SIC.
Here, $\beta_{Q_{2}}$ is obtained to satisfy QoS requirement at Ru$_2$ without using SIC as
$R_{_{\mathrm{Ru}_2}}^\mathrm{NSIC}(\beta_{Q_{2}})=Q$.

For $Q\leq r_{2}$, since if $\beta_{0}>\tilde{\beta}_{1}> \beta_{Q_{2}}$, the optimal $\beta$
to guarantee QoS requirement at Ru$_2$ is given by $\beta^\star=\tilde{\beta}_{1}$, the condition that makes
$\beta^\star=\tilde{\beta}_{1}$ is obtained as
\begin{align}\left\{\begin{array}{ll}
\tilde{\beta}_{1}< \beta_{0}&\Rightarrow~g_{0}> g_{12},\\
\tilde{\beta}_{1}> \beta_{Q_{2}}&\Rightarrow~Q> r_3,
\end{array}\right.
\end{align}
where $r_3$ is given in \eqref{rs}.
Therefore, if $g_{0}> g_{12}\geq g_{1}$ and $r_{2}\geq Q> r_{3}$,
the optimal $\beta$ is obtained by $\beta^\star=\tilde{\beta}_{1}$ and QoS requirement at Ru$_2$ is achieved by applying SIC at Ru$_2$.

Otherwise, the optimal $\beta$ is obtained by $\beta^\star=\beta_{Q_2}$ and QoS requirement at Ru$_2$ is achieved without applying SIC at Ru$_2$.

\section{Proof of Theorem \ref{theorem2}}\label{app2}
\setcounter{equation}{0}
\renewcommand{\theequation}{B.\arabic{equation}}

For given $\beta$, by substituting $\bw_1^{\mathrm{opt}}$ of Lemma \ref{lemma_ryu} into \eqref{R1-MISO-app}, $\tilde{R}_{_{\mathrm{Ru}_1}}\!(\bw_1)$ can be rewritten by
\begin{align}\label{R1-MISO-app1}
\tilde{R}_{_{\mathrm{Ru}_1}}(\eta)=&\frac{\alpha}{2}\log_2\left\{\rho_1\min\Big(g(\eta)+m_1(\beta),
~f(\eta)\Big)\right\}\nonumber\\
&+\frac{1-\alpha}{2}\log_2\left(\rho_1g(\eta)\right),
\end{align}
where $g(\eta)$ and $f(\eta)$ are given by
\begin{align}
g(\eta)&\triangleq(\sqrt{\eta v_1}+\sqrt{(1-\eta)v_2})^2=|\bh_1^\HH\bw_1|^2,\\
f(\eta)&\triangleq\eta \tilde{g}_{0}=|\bh_0^\HH\bw_1|^2.
\end{align}

First, if $\tilde{g}_{0}$ is large as $\tilde{g}_{0}\!\!>\!\!\frac{(v_1+v_2)^2}{v_1}$ and
power allocation for cooperation, $\beta$, is small as
$\beta\!\!<\!\!\underline{\beta}$, we have $\!f(\eta)\!\!>\!\!g(\eta)\!\!+\!\!m_1(\beta)\!$ for any $\eta$ in $\frac{v_1}{v_1\!+\!v_2}\!\leq\!\eta\!\leq\!1$ and thus, $\tilde{R}_{_{\mathrm{Ru}_1}}\!(\eta)$ in \eqref{R1-MISO-app1} can be given as
\begin{align}\label{R1-g}
\!\!\!\!\!\tilde{R}_{_{\mathrm{Ru}_1}}\!(\eta)\!\!=\!\!\frac{\alpha}{2}\!\log_2\!\left\{\!\rho_1\Big(g(\eta)
\!\!+\!\!m_1(\beta)\!\Big)\!\right\}\!\!+\!\!\frac{1\!\!-\!\!\alpha}{2}\!\log_2\!\left(\!\rho_1g(\eta)\right)\!.\!\!\!\!\!\!
\end{align}
Since \eqref{R1-g} is a decreasing function with $\eta$ in $\frac{v_1}{v_1+v_2}\leq\eta\leq1$, we can obtain $\eta^\star$ that maximize \eqref{R1-g} as $\eta^\star=\frac{v_1}{v_1+v_2}$.

Contrarily, if $\tilde{g}_{0}$ is small as $\tilde{g}_{0}<v_{1}$ or $\beta$ is large as $\beta>\overline{\beta}$
for $\tilde{g}_{0}\geq v_{1}$,
we have $f(\eta)<g(\eta)+m_1(\beta)$ for any $\eta$ in $\frac{v_1}{v_1+v_2}\leq\eta\leq1$. In this case, \eqref{R1-MISO-app1} can be rewritten as
\begin{equation}\label{R1-f}
\tilde{R}_{_{\mathrm{Ru}_1}}(\eta)=\frac{\alpha}{2}\log_2\left(\rho_1f(\eta)\right)+\frac{1-\alpha}{2}\log_2\left(\rho_1g(\eta)\right).
\end{equation}
Since \eqref{R1-f} is a concave function with respect to $\eta$, we obtain $\eta^\star$ to maximize \eqref{R1-f} by solving $\frac{\partial\tilde{R}_{_{\mathrm{Ru}_1}}(\eta)}{\partial\eta}=0$ as $\eta^\star=\eta_2$, which is given in \eqref{eta}.

Otherwise, according to $\beta$,
$\tilde{R}_{_{\mathrm{Ru}_1}}(\eta)$ in \eqref{R1-MISO-app1} can be represented by either \eqref{R1-g} or \eqref{R1-f}. We first derive $\eta_3(\beta)$, which is given in \eqref{eta3}, to satisfy $f(\eta_3(\beta))=g(\eta_3(\beta))+m_1(\beta)$. Then, for $\eta_3(\beta)\leq\eta\leq1$, $\tilde{R}_{_{\mathrm{Ru}_1}}(\eta)$ is represented by \eqref{R1-g}, which is a decreasing function of $\eta$ and thus, we can obtain $\eta^\star=\eta_3(\beta)$, given in \eqref{eta}. For $\frac{v_1}{v_1+v_2}\leq\eta\leq\eta_3(\beta)$, $\tilde{R}_{_{\mathrm{Ru}_1}}(\eta)$ is represented by \eqref{R1-f}, which is a concave function of $\eta$ achieving maximum value at $\eta^\star=\eta_2$. Therefore, if $\eta_2$ is in the range of $\frac{v_1}{v_1+v_2}\leq\eta\leq\eta_3(\beta)$, we can obtain $\eta^\star=\eta_2$. Otherwise, if $\eta_2>\eta_3(\beta)$, we obtain $\eta^\star=\eta_3(\beta)$ because \eqref{R1-f} is an increasing function of $\eta$ for $\frac{v_1}{v_1+v_2}\leq\eta\leq\eta_3(\beta)$.
Consequently, if $\eta_2\leq \eta_{3}(\beta)$, we obtain the optimal $\eta$ as $\eta^\star=\eta_2$,
and otherwise, the optimal $\eta$ is obtained by $\eta^\star=\eta_3(\beta)$. Therefore, the optimal $\eta$ can be represented by
$\eta^\star=\min\{\eta_2,\eta_3(\beta)\}$.

\bibliography{Reference}

\begin{IEEEbiography}[{\includegraphics[width=1in,height=1.25in,keepaspectratio]{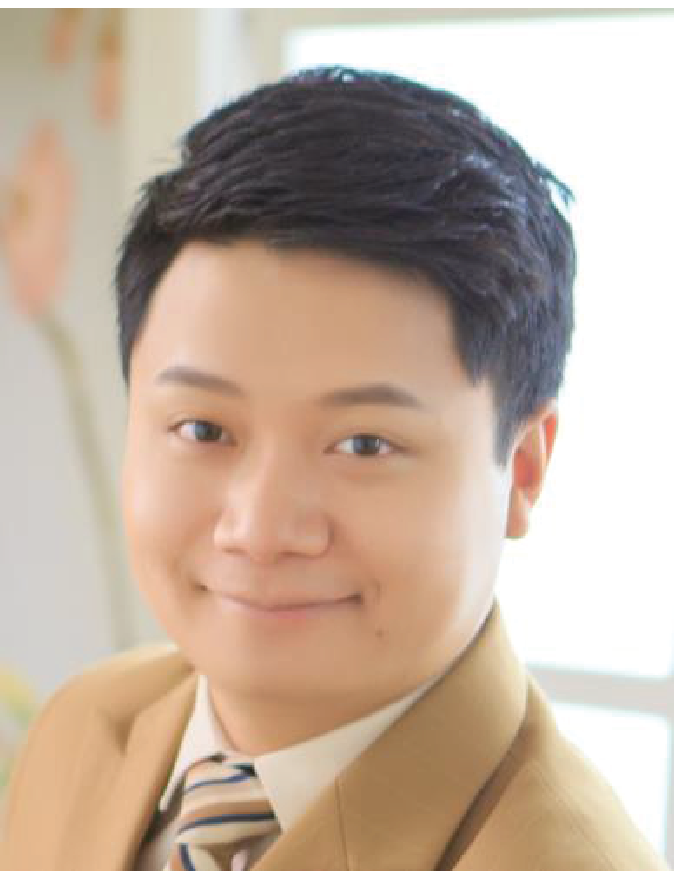}}]
{Mingxiong Zhao} received the B.S. degree in Electrical Engineering and the Ph.D. degree in Information and Communication Engineering from South China University of Technology (SCUT), Guangzhou, China, in 2011 and 2016, respectively. He was a visiting Ph.D. student at University of Minnesota (UMN), Twin Cities, MN, USA, from 2012 to 2013 and Singapore University of Technology and Design (SUTD), Singapore, from 2015 to 2016, respectively. Since 2016, he has been an Assistant Professor at the School of Software, Yunnan University, Kunming, China. His current research interests are physical layer security, cooperative relay communication, and social aware communication systems.
\end{IEEEbiography}

\begin{IEEEbiography}[{\includegraphics[width=1in,height=1.25in,keepaspectratio]{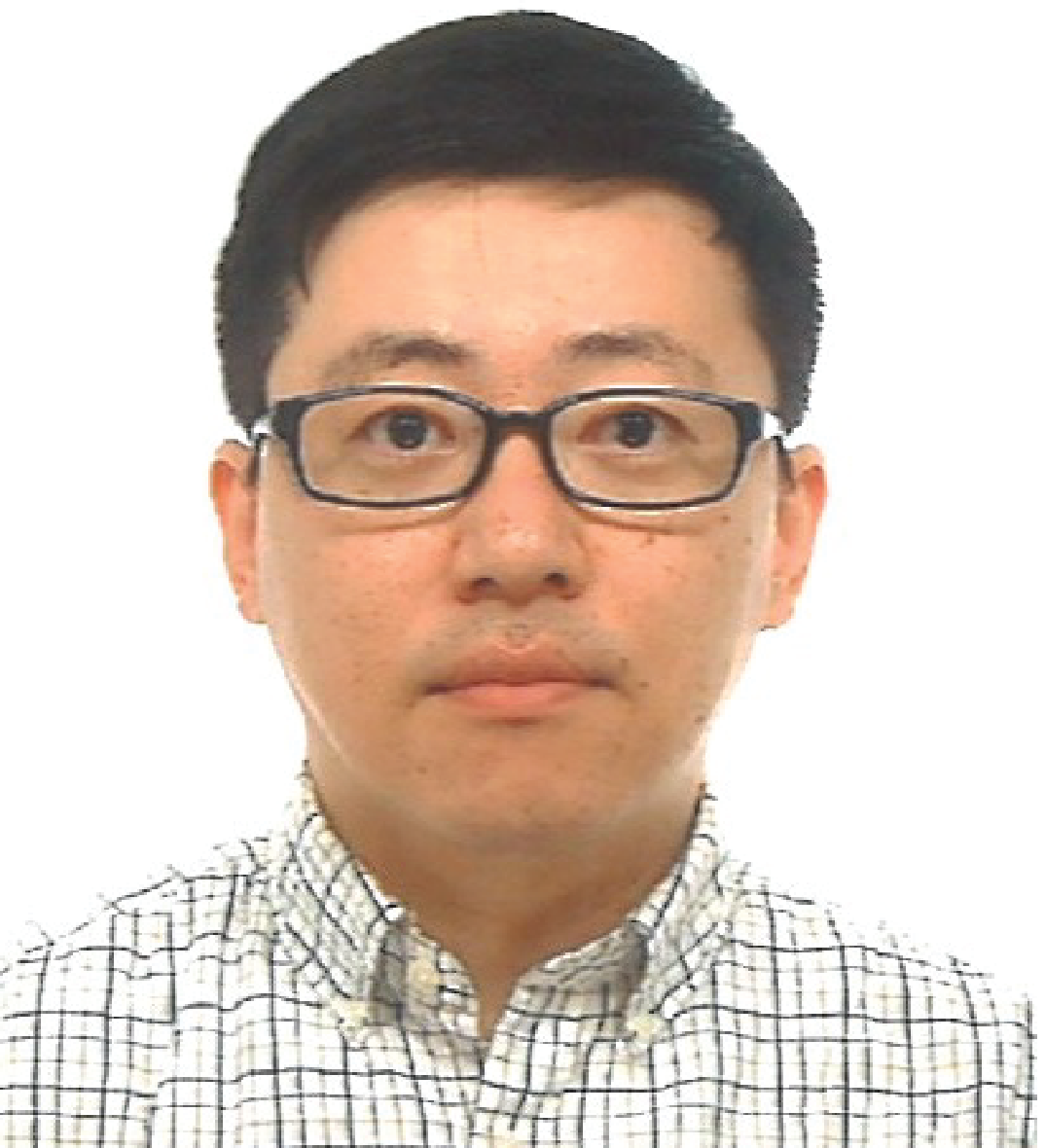}}]
{Jong Yeol Ryu}(S'11-M'14) received the B.E.\ degree in Electrical Engineering from Chungnam National University, Daejeon, Korea, in 2008. He received the M.S.\ and Ph.D.\ degrees in Electrical Engineering from Korea Advance Institute of Science and Technology (KAIST), Daejeon, Korea, in 2010 and 2014, respectively. From 2014 to 2016, he was a Postdoctoral Fellow at Singapore University of Technology and Design (SUTD). Currently, he is an Assistant Professor at the Department of Information and Communication Engineering, Gyeongsang National University (GNU), Tongyeong, Korea. His current research interests are communication secrecy, cooperative user relay communications, and social aware communication systems.
\end{IEEEbiography}

\begin{IEEEbiography}[{\includegraphics[width=1in,height=1.25in,clip,keepaspectratio]{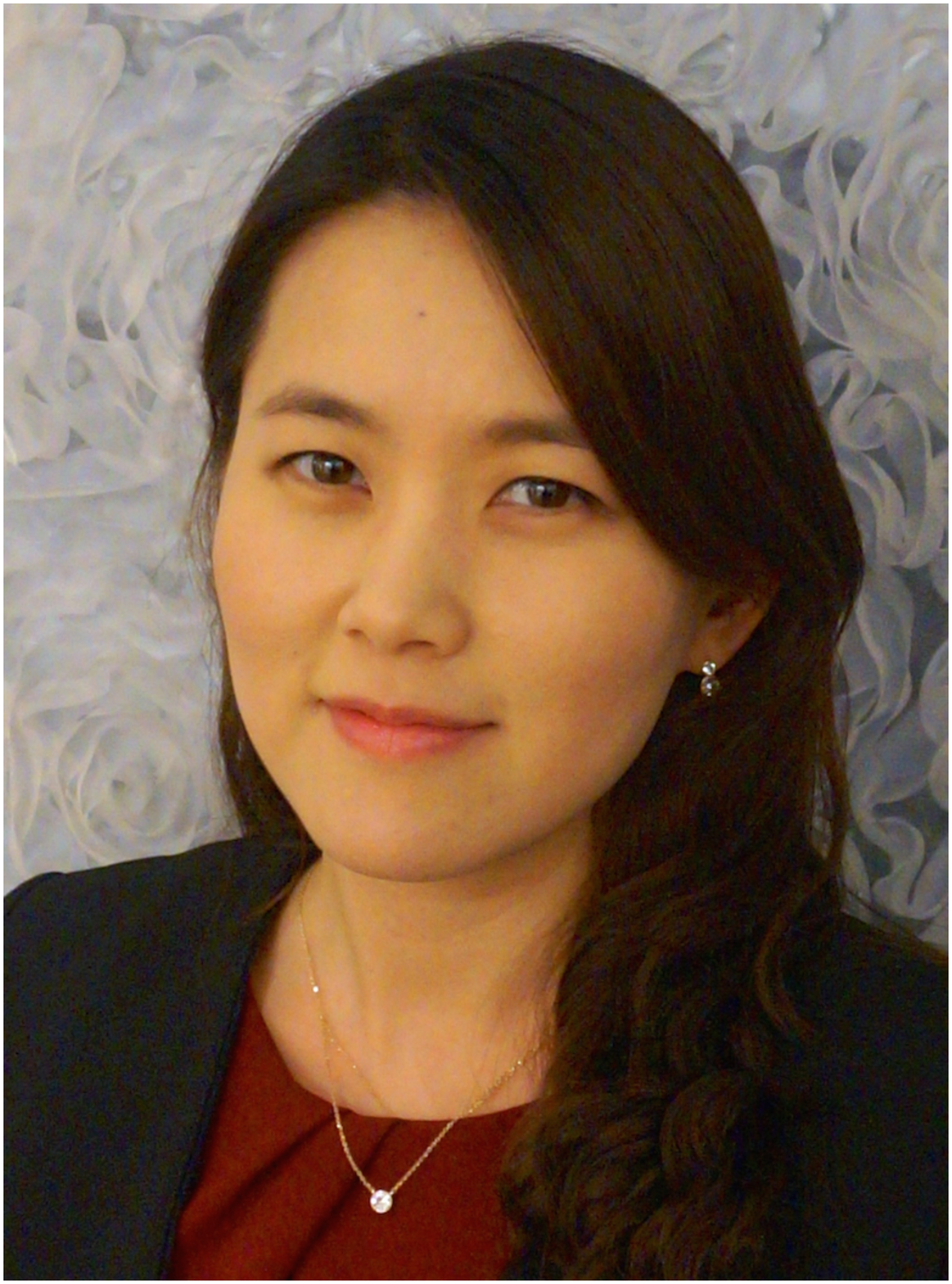}}]
{\bfseries Jemin Lee} (S'06-M'11) received the B.S. (with high honors), M.S., and Ph.D. degrees in Electrical and Electronic Engineering from Yonsei University, Seoul, Korea, in 2004, 2007, and 2010, respectively.
She was a Postdoctoral Fellow at the Massachusetts Institute of Technology (MIT), Cambridge, MA from 2010 to 2013,
and a Temasek Research Fellow at iTrust, Centre for Research in Cyber Security, Singapore University of Technology and Design (SUTD), Singapore from 2014 to 2016. Currently, she is an Assistant Professor at the Department of Information and Communication Engineering, Daegu Gyeongbuk Institute of Science and Technology (DGIST), Daegu, Korea. Her current research interests include physical layer security, wireless security, heterogeneous networks, and machine-type communication.

Dr.~Lee is currently an Editor for the {\scshape IEEE Transactions on Wireless Communications} and the {\scshape IEEE Communications Letters}, and served as a Guest Editor for the {\scshape IEEE Wireless Communications}, special issue on LTE in Unlicensed Spectrum, 2016, and the {\scshape ELSEVIER Physical Communication}, special issues on Physical Layer Security in 2016 and Heterogeneous and Small Cell Networks in 2014. She received the IEEE ComSoc Asia-Pacific Outstanding Young Researcher Award in 2014, the Temasek Research Fellowship in 2013, the Chun-Gang Outstanding Research Award in 2011, and the IEEE WCSP Best Paper Award in 2014.
\end{IEEEbiography}

\begin{IEEEbiography}[{\includegraphics[width=1in,height=1.25in,keepaspectratio]{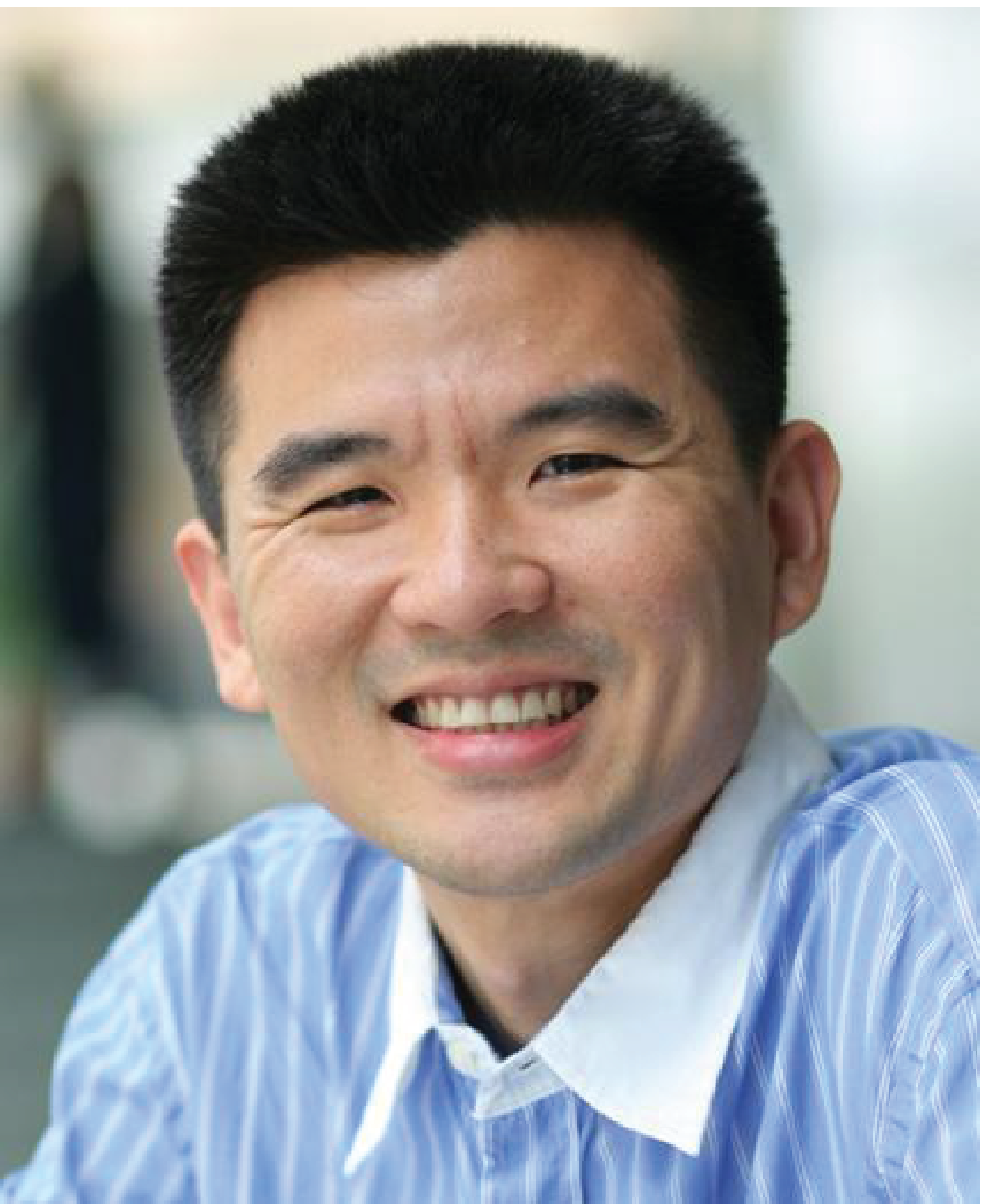}}]
{Tony Q.S. Quek}(S'98-M'08-SM'12) received the B.E.\ and M.E.\ degrees in Electrical and Electronics Engineering from Tokyo Institute of Technology, respectively. At MIT, he earned the Ph.D.\ in Electrical Engineering and Computer Science. Currently, he is a tenured Associate Professor with the Singapore University of Technology and Design (SUTD). He also serves as the Associate Head of ISTD Pillar and the Deputy Director of the SUTD-ZJU IDEA. His main research interests are the application of mathematical, optimization, and statistical theories to communication, networking, signal processing, and resource allocation problems. Specific current research topics include heterogeneous networks, wireless security, internet-of-things, and big data processing.

Dr.\ Quek has been actively involved in organizing and chairing sessions, and has served as a member of the Technical Program Committee as well as symposium chairs in a number of international conferences. He is serving as the Workshop Chair for IEEE Globecom in 2017, the Tutorial Chair for the IEEE ICCC in 2017, and the Special Session Chair for IEEE SPAWC in 2017. He is currently an elected member of IEEE Signal Processing Society SPCOM Technical Committee. He was an Executive Editorial Committee Member for the {\scshape IEEE Transactions on Wireless Communications}, an Editor for the {\scshape IEEE Transactions on Communications}, and an Editor for the {\scshape IEEE Wireless Communications Letters}. He is a co-author of the book ``Small Cell Networks: Deployment, PHY Techniques, and Resource Allocation" published by Cambridge University Press in 2013 and the book ``Cloud Radio Access Networks: Principles, Technologies, and Applications" by Cambridge University Press in 2017.

Dr.\ Quek was honored with the 2008 Philip Yeo Prize for Outstanding Achievement in Research, the IEEE Globecom 2010 Best Paper Award, the 2012 IEEE William R. Bennett Prize, the IEEE SPAWC 2013 Best Student Paper Award, the IEEE WCSP 2014 Best Paper Award, the 2015 SUTD Outstanding Education Awards -- Excellence in Research, the 2016 Thomson Reuters Highly Cited Researcher, and the 2016 IEEE Signal Processing Society Young Author Best Paper Award.
\end{IEEEbiography}

\begin{IEEEbiography}[{\includegraphics[width=1in,height=1.25in,keepaspectratio]{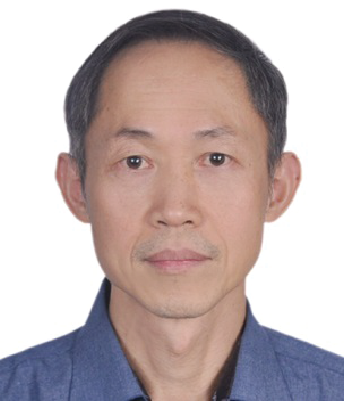}}]
{Suili Feng} (M'05) received the B.S. degree in Electrical Engineering and the M.S. and Ph.D. degrees in Electronic and Communication System from South China University of Technology, Guangzhou, China, in 1982, 1989, and 1998, respectively. He was a Research Assistant with Hong Kong Polytechnic University, Hong Kong, from 1991 to 1992, and a Visiting Scholar with University of South Florida, Tampa, FL, USA, from 1998 to 1999. He has been with South China University of Technology since 1989, where he is currently a Professor with the School of Electronic and Information Engineering. His research interests include wireless networks, computer networks, and communication signal processing.
\end{IEEEbiography}

\end{document}